\documentclass[9pt]{article}

\usepackage[utf8]{inputenc}   
\usepackage{amsmath, amssymb, amsfonts} 
\usepackage{graphicx}         
\usepackage{hyperref}         
\usepackage{geometry}         
\usepackage{caption}          
\usepackage{cite}             
\usepackage{authblk}          

\usepackage[linguistics]{forest}
\usepackage{xspace}
\usepackage{algorithm}     
\usepackage[noend]{algpseudocode} 

\newcommand{\myitem}[1]{\refstepcounter{enumi}\item[$($#1.\theenumi$)$]}
\newcommand{\mygraphsize}{0.29}
\newcommand{\libname}{TRAM\xspace}
\newcommand{\circlednum}[1]{\tikz[baseline=(char.base)]\node[shape=circle, draw=black, fill=black, text=white, inner sep=0.8pt] (char) {\scriptsize\textbf{#1}};%
}

\title{Scaling atomic ordering in shared memory}
\author[1]{Lorenzo Martignetti}
\author[1]{Eli\~a Batista}
\author[2]{Gianpaolo Cugola}
\author[1]{Fernando Pedone}
\affil[1]{Università della Svizzera Italiana, Lugano, Switzerland}
\affil[2]{Politecnico di Milano, Milan, Italy}
\date{\today}

\begin{document}

\maketitle

\begin{abstract}
  Atomic multicast is a communication primitive used in dependable systems to ensure consistent ordering of messages delivered to a set of replica groups. This primitive enables critical services to integrate replication and sharding (i.e., state partitioning) to achieve fault tolerance and scalability. While several atomic multicast protocols have been developed for message-passing systems, only a few are designed for the shared memory system model. This paper introduces \libname, an atomic multicast protocol specifically designed for shared memory systems, leveraging an overlay tree architecture. Due to its simple and practical design, \libname delivers exceptional performance, increasing throughput by more than 3$\times$ and reducing latency by more than 2.3$\times$ compared to state-of-the-art shared memory-based protocols. Additionally, it significantly outperforms message-passing-based protocols, boosting throughput by up to 5.9$\times$ and reducing latency by up to 106$\times$.
\end{abstract}

\section{Introduction}

\begin{figure*}
	\center
	\begin{forest}
		[\bf Consistent order across\\\bf shards \& replicas
		[message passing
		[Atomic broadcast\\(single shard) [Ring Paxos \cite{marandi2010ring}\\Raft$^*$ \cite{184040}\\S-Paxos$^*$ \cite{biely2012s}]]
		[Atomic multicast\\(multiple shards)
		[genuine [FastCast \cite{FastCast17}\\PrimCast \cite{Pacheco2023}\\WhiteBox \cite{gotsman2019white}\\Spanner$^*$\cite{CDE12}]]
		[partially\\genuine, base=middle [ByzCast \cite{ByzCast}\\Multi-Ring\\Paxos \cite{marandi2012multi}]]]]
		[\bf shared memory
		[Atomic broadcast\\(single shard) [APUS$^*$ \cite{APUS}\\DARE$^*$ \cite{DARE}\\Mu$^*$ \cite{Mu}\\P4CE$^*$ \cite{P4CE}]]
		[\bf Atomic multicast\\\bf (multiple shards)
		[genuine
			[Derecho \cite{jha2019derecho}\\RamCast \cite{le2021ramcast}\\FaRM$^*$ \cite{FaRM}]]
		[\bf partially\\\bf genuine, base=middle
		[\bf \libname\\\bf (this paper), draw]]]]]
	\end{forest}
	\captionsetup{font=small}
	\caption{Message-passing versus shared memory (RDMA) protocols that ensure consistent order across shards (partitions) and replicas; starred protocols ($^*$) do not encapsulate ordering logic in atomic broadcast and atomic multicast primitives; see Section \ref{sec:related-work} for more details.}
	\label{fig:allsystems}
\end{figure*}
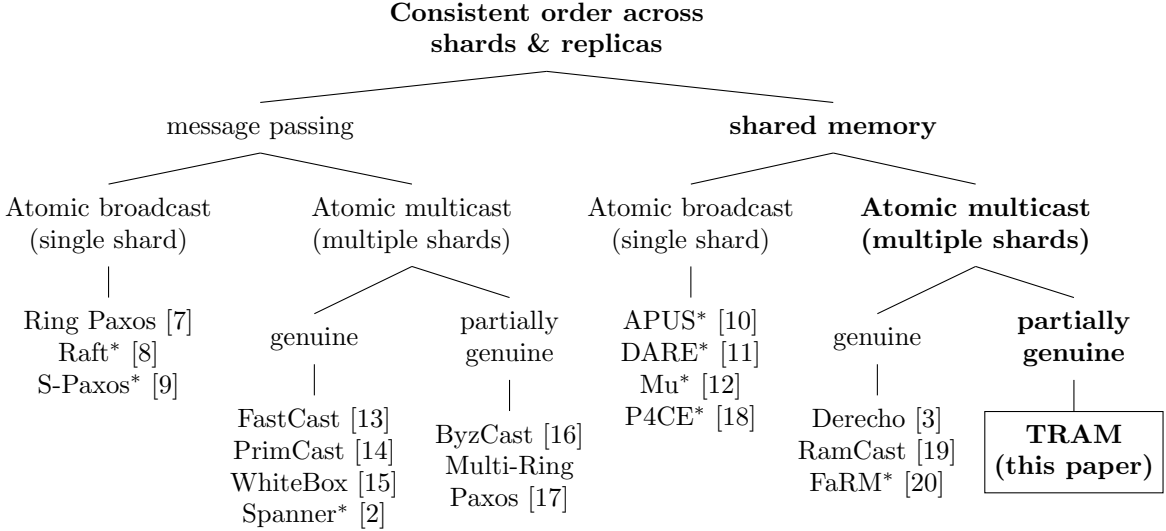

Many critical services combine replication and sharding to achieve fault tolerance and scalability.
In general, the application state is partitioned into shards, and each shard is fully replicated by a group of replicas (e.g., \cite{Long2019, CDE12, jha2019derecho, thomson2012calvin}).
To support this model while ensuring strong consistency (e.g., linearizability \cite{herlihy1990linearizability}, serializability \cite{Pap86}), replicas must order client requests within and across replica groups.
Several protocols have been proposed to order requests in such environments based on message passing and shared memory communication (see Figure~\ref{fig:allsystems}).

\emph{Message-passing vs. shared-memory communication.}
For years, practical distributed systems have been developed based on message-passing communication (e.g., \cite{marandi2010ring, 184040, biely2012s, CDE12}).
Recent advances in shared-memory technology, however, have enabled systems to benefit from improved communication (e.g., \cite{APUS, DARE, Mu, jha2019derecho}).
Remote Direct Memory Access (RDMA) is a remarkable technology that provides high-throughput and low-latency communication by implementing network stack layers in hardware.
Additionally, processes can access remote memories without involving the remote CPU.
Compared to message-passing-based systems, RDMA introduces memory management and synchronization challenges:
servers must explicitly manage memory buffers, and multiple servers might access a memory region concurrently, creating race conditions.

\emph{Atomic broadcast vs. atomic multicast.}
Some replication protocols explicitly encapsulate the logic used to order requests using group communication primitives.
Two well-established abstractions are atomic broadcast and atomic multicast \cite{HT93}.
With atomic broadcast, client requests can be reliably and consistently propagated to all system replicas, inducing a total order on requests (e.g., \cite{marandi2010ring, 184040, APUS, P4CE}).
Atomic multicast is a group communication abstraction that generalizes atomic broadcast by partitioning the set of replicas into multiple groups (i.e., shards) and allowing clients to send each message to a subset of all the available groups, with reliability and order guarantees (e.g., \cite{FastCast17, gotsman2019white, ByzCast, jha2019derecho, FaRM}).
Intuitively, atomic multicast guarantees that all non-faulty replicas addressed by a request deliver the request, and the replicas agree on the order in which the requests are delivered.
Atomic multicast provides strong communication guarantees and should not be confused with ``best-effort'' network-level communication primitives (e.g., IP multicast).

\emph{Genuine vs. non-genuine atomic multicast.}
Messages multicast to different sets of destination groups may interleave in non-obvious ways, which renders implementing message order in a distributed system challenging.
Atomic multicast algorithms can be classified according to which replicas coordinate to order messages into genuine, partially genuine, or non-genuine.
While in a \emph{genuine} algorithm only the message sender and replicas in destination groups coordinate to propagate and order a multicast message~\cite{GS01b}, in a \emph{non-genuine} algorithm, any groups of replicas may be involved.
For example, message ordering can be delegated to a fixed group of replicas, regardless of the message's destination.
A \emph{partially genuine} atomic multicast protocol ensures that messages multicast to a single group of replicas are ordered within the destination group.
An important class of partially genuine atomic multicast protocols assumes that groups communicate using an overlay tree to order messages (e.g., \cite{ByzCast}).
The main distinguishing feature of a tree-based atomic multicast protocol is its relative simplicity compared to genuine protocols.
In general, a simple protocol has the advantage of being more amenable to performance optimizations than a complex protocol.

\emph{This paper's contributions.}
This paper introduces \libname, a tree-based atomic multicast protocol designed explicitly for the shared-memory communication model. Shared memory systems enable efficient communication through direct memory access, reducing CPU utilization and communication overhead while allowing granular control over memory access permissions. By revoking write permissions before modifying shared memory, a process can ensure safe, isolated writes without additional synchronization.
\libname leverages these capabilities to achieve significant performance improvements using an overlay tree that balances communication and synchronization costs while maintaining strong consistency.
It increases throughput by more than 3$\times$ and reduces latency by more than 2.3$\times$ compared to state-of-the-art shared memory-based protocols and outperforms message-passing-based protocols, boosting throughput by up to 5.9$\times$ and reducing latency by up to 106$\times$.
\libname's partially genuine design reduces coordination overhead among processes and optimizes shared memory operations, demonstrating its effectiveness in addressing fault tolerance, scalability, and consistent message ordering in distributed systems.

\emph{Paper organization.}
The remainder of this paper is structured as follows:
Section~\ref{sec:background} provides the background and system model necessary for understanding \libname.
Section~\ref{sec:design:motiv} showcases the motivations behind \libname design, which is explained in Section~\ref{sec:design}.
Section~\ref{sec:algodetails} details the algorithm, including data structures, client operations, and replica coordination.
Section~\ref{sec:proof} proves \libname correctness, while Section~\ref{sec:implementation} describes the implementation.
Section~\ref{sec:perfeval} evaluates \libname's performance under various configurations and workloads, comparing it to state-of-the-art protocols.
Section~\ref{sec:related-work} highlights related work, and Section~\ref{sec:conclusions} concludes the paper and outlines potential directions for future research.
\section{Background}
\label{sec:background}


\subsection{System model}
\label{sec:system-model}

We assume a distributed system model in which processes use shared-memory communication \cite{Aguilera2019}.
The system is composed of a set of client and server (replica) processes.
Processes communicate by accessing portions of each other's memory.
Processes are \emph{correct} if they do not fail, or \emph{faulty}, otherwise.
In either case, processes do not experience arbitrary behavior (i.e., no Byzantine failures).


The system is asynchronous: there are no fixed upper bounds on communication and relative process speeds.
To order messages within a group \cite{FLP85,CT96}, we assume that replicas at each group have access to a weak leader election oracle~\cite{CT96}, defined as follows.
For any group $g$, the oracle outputs a single process such that there is (a)~a correct process $p \in g$ and (b)~a time after which, every invocation of $\texttt{leader(}g\texttt{)}$ returns $p$.

%
%

A process can share memory regions with other processes and define permissions for those shared memory regions.
Process $q$ can read and write $p$'s memory region $mr$ with remote operations $\texttt{read}(p,mr,value)$ and $\texttt{write}(p,mr,value)$, respectively.
A permission associated with memory region $mr$ defines disjoint sets of processes, $R_{mr}, W_{mr}$, and $RW_{mr}$, that can read, write, and read-write the memory region $mr$.
Process $q$ has permission to read (respectively, write and read-write) $p$'s $mr$ if $q \in R_{mr}$ (resp., $W_{mr}, RW_{mr}$).
A process can initially assign permissions for its shared memory regions and later change these permissions.
If both $p$ and $q$ are correct processes and $p$ has permission to access $q$'s memory region, then any read or write operation performed by $p$ on that region will succeed.


\subsection{Atomic multicast properties}
\label{sec:amcast}

Let $\Pi$ be the set of server processes in the system and $\Gamma \in 2^{\Pi}$ the set of process groups, where $|\Gamma| = k$.
Groups are disjoint and each group contains $n = 2f+1$ processes, where $f$ is the maximum number of faulty processes per group.
The assumption about disjoint groups has little practical implication since it does not prevent collocating processes that are members of different groups on the same machine.
Yet, it is important since atomic multicast requires stronger assumptions when groups intersect~\cite{GS01b}.
A set of $f + 1$ processes in group $g$ is a \emph{quorum} in $g$.

A process multicasts a message $m$ with unique nonnull identifier $m.id$ to groups in $m.dst$ by invoking primitive $\texttt{multicast}(m)$, where $m.dst$ is a special field in $m$ specifying its destination groups. A process delivers $m$ with primitive $\texttt{deliver}(m)$.
We define the relation $<$ on the set of messages delivered by processes as follows: $m < m'$ iff there exists a process that delivers $m$ before $m'$.

Atomic multicast ensures the following properties:
\begin{itemize}
  \item \textit{Validity}:~if a correct process $p$ multicasts a message $m$, then eventually all correct processes $q \in g$, where $g \in m.\mathit{dst}$, deliver $m$.
  \item \textit{Uniform agreement}:~if a process $p$ delivers a message $m$, eventually all correct processes $q \in g$, where $g \in m.\mathit{dst}$, deliver $m$.
  \item \textit{Uniform integrity}:~for any process $p$ and any message $m$, $p$ delivers $m$ at most once, and only if $p \in g$, $g \in m.\mathit{dst}$, and $m$ was previously multicast.
  \item \textit{Uniform prefix order}:~for any messages $m$ and $m'$ and processes $p$ and $q$ such that $p \in g$, $q \in h$ and $\{ g, h \} \subseteq m.\mathit{dst} \cap m'.\mathit{dst}$, if $p$ delivers $m$ and $q$ delivers $m'$, either $p$ delivers $m'$ before $m$ or $q$ delivers $m$ before $m'$.
  \item \textit{Uniform acyclic order}:~the relation $<$ is acyclic.
\end{itemize}
Atomic broadcast is a special case of atomic multicast in which there is a single group (i.e., $\Gamma$ is a singleton).

\subsection{Classes of atomic multicast protocols}

An atomic multicast algorithm is \emph{genuine}~\cite{GS01b} if for any run of the algorithm and any process $p$, if $p$ participates in the run, then a message $m$ is multicast in the run and either $p$ is a destination or sender of $m$.
Process $p$ participates in a run if $p$ sends or receives messages, or executes a remote read or write operation in the run.

In a \emph{partially genuine} atomic multicast protocol~\cite{ByzCast}, processes that are neither the destination nor sender of a multicast message can take part in the run only if the message is addressed to more than one group.
More precisely, for any run of the algorithm and any process $p$, if $p$ participates in the run, then a message $m$ is multicast in the run, and (i) $p$ is a destination or sender of $m$; or (ii) $m$ is addressed to more than one group.

\subsection{Remote Direct Memory Access}
\label{sec:rdma}

Remote Direct Memory Access (RDMA) is a technology that enables direct data access to a remote machine's memory without involving the operating system and processor of the remote machine.
RDMA implements the network stack in hardware and provides both low-latency and high-bandwidth communication by bypassing the kernel and supporting zero-copy communication.
RDMA provides two-sided operations (e.g., send, receive), one-sided operations (e.g., read, write), and atomic operations (e.g., compare-and-swap, fetch-and-increment). The two-sided operations involve the CPU of the remote host and rely on user-space memory copies.
Thus, they introduce overhead when compared to one-sided RDMA verbs \cite{FaRM}.
Besides, previous studies have established that remote write operations provide performance superior to both remote reads and send and receive operations, and much better performance than atomic operations \cite{kalia2014using, kalia2016design, mitchell2013using}.

RDMA provides three transport modes: Reliable Connection (RC), Unreliable Connection (UC), and Unreliable Datagram (UD).
While RC and UC are connection-oriented and support only one-to-one data transmission, UD supports one-to-one and one-to-many transmission without establishing connections.
RC ensures data transmission is reliable and correct at the network layer, while UC does not have such a guarantee.
In this work, we use RC to provide in-order reliable delivery.

To establish a connection between two remote hosts, the RDMA-enabled network card (RNIC) on each host creates a logical RDMA endpoint known as a Queue Pair (QP), including a send queue and a receive queue for storing data transfer requests.
Operations are posted to QPs as Work Requests (WRs) to be consumed and executed by the RNIC.
When a RDMA operation is completed, a completion event is pushed to a Completion Queue (CQ).
Each host makes local memory regions (MR) available for remote access by asking its operating system to pin the memory pages that would be used by the RNIC.

Both QPs and MRs can have different access modes (i.e., read-only, write-only or read-write).
The hosts specify the access mode when initializing the QP or registering the MR, but the access mode can be dynamically updated later.
The host can register the same memory for different MRs. Each MR then has its own access mode.
In this way, different remote machines can have different access rights to the same memory region.

We assume that reads and writes of shared variables are atomic, implemented using a canary approach \cite{APUS, Mu} (see Section \ref{sec:implementation} for a detailed discussion).

\section{Motivation}
\label{sec:design:motiv}

Most atomic multicast algorithms aim to reduce the number of communication steps (i.e., network delays in the critical path) and achieve genuineness (e.g., \cite{FastCast17, Pacheco2023, gotsman2019white, le2021ramcast, GS01b}).
While this design is effective in message-passing environments, minimizing processing and communication overhead, this section argues that shared-memory systems benefit from different optimization strategies.

\begin{figure}[htbp]
  \centering
  \includegraphics[width=0.4\textwidth]{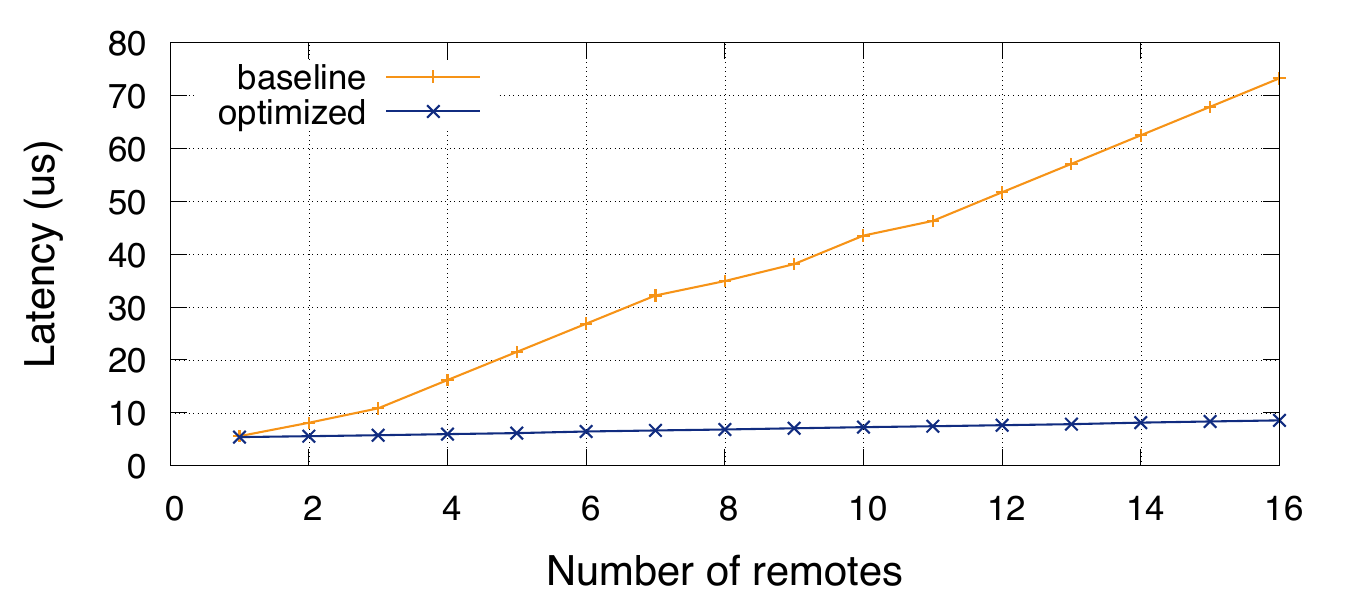}
  \caption{ Average latency of remote write operations as a function of the number of target processes. In the baseline configuration, the initiator waits for the completion of each write operation before issuing the next, leading to a linear increase in latency with the number of targets. In the optimized configuration, all write operations are issued concurrently, and completions are awaited collectively, significantly reducing latency. \libname leverages this batched completion strategy to minimize remote write overhead.}
  \label{fig:write-lat}
\end{figure}

\begin{enumerate}
  \item First, in message-passing protocols, a process typically directly communicates with processes in any destination group. In shared memory environments, this often requires scanning input buffers proportional in size to the number of potential senders (e.g., \cite{Mu, FaRM, le2021ramcast}), which can lead to performance degradation. Although this overhead is manageable in message-passing systems, it becomes significant in RDMA-based shared-memory systems where operations occur at a microsecond scale.
  \item Second, RDMA permissions can be leveraged to reduce process coordination. Once permissions are established, the success or failure of a remote read or write inherently conveys information about the remote process's willingness to accept the operation. As RDMA operations bypass the remote CPU, processes can inspect remote state without involving the remote processor, reducing synchronization overhead.
  \item Third, remote writes are the most efficient RDMA communication primitive \cite{le2021ramcast, Kalia2014, mitchell2013using}, but their cost, including completion polling, is not negligible. Although RDMA allows issuing multiple remote operations concurrently and awaiting their completion collectively, serializing writes is sometimes unavoidable, particularly when the outcome of one operation determines subsequent actions. Figure \ref{fig:write-lat} illustrates how remote write latency scales with the number of targets.
\end{enumerate}

In shared-memory environments, optimizing solely for the number of communication steps, a common goal in message-passing atomic multicast protocols, is insufficient. Factors like local buffer access, RDMA permission handling, and remote operation batching critically affect performance.

In response to these considerations, this paper introduces TRAM, a partially genuine, tree-based protocol that balances communication and synchronization costs. Rather than coordinating across all destination groups, each group interacts only with its parent and children in a static overlay tree. Within each group, RDMA permissions allow leaders to impose order by directly writing to followers' logs, eliminating the need for coordination among replicas in failure-free executions. The tree overlay also enables parallelism by allowing different subtrees to proceed independently, thus exploiting RDMA's parallel write capabilities.
\section{\libname overview}
\label{sec:design}

\libname assumes that communication among groups follows a predefined tree overlay $\mathcal{T}$, known to clients and replicas in advance.
This paper considers a static overlay, meaning the structure of $\mathcal{T}$ remains fixed throughout the execution.
Relaxing this assumption to support dynamic overlays and adapt to evolving communication patterns is a direction for future work. This could potentially leverage techniques proposed in prior research \cite{flexcast}.

To explain the algorithm, consider the following definitions:
\begin{itemize}
  \item the \emph{least common ancestor} (\emph{lca}) of a set of destination groups is the closest group to $\mathcal{T}$'s root from which all destinations can be reached traversing $\mathcal{T}$ downwards; and
  \item the \emph{reach} of a group $g$ in $\mathcal{T}$ is the set of groups reachable from $g$ traversing the tree downwards, $g$ included.
\end{itemize}


Clients and replicas communicate using shared buffers \cite{FaRM}.
Within a group, \libname leverages RDMA permissions to efficiently order messages within the group.
Each group has a leader who proposes the order of messages to the passive followers by writing in their log remotely.
If the leader fails, a follower takes over.
The protocol tolerates a minority of process failures within the group and allows multiple followers to attempt replacing a crashed leader.


\begin{figure}[htbp]
  \centering
  \includegraphics[width=0.4\textwidth]{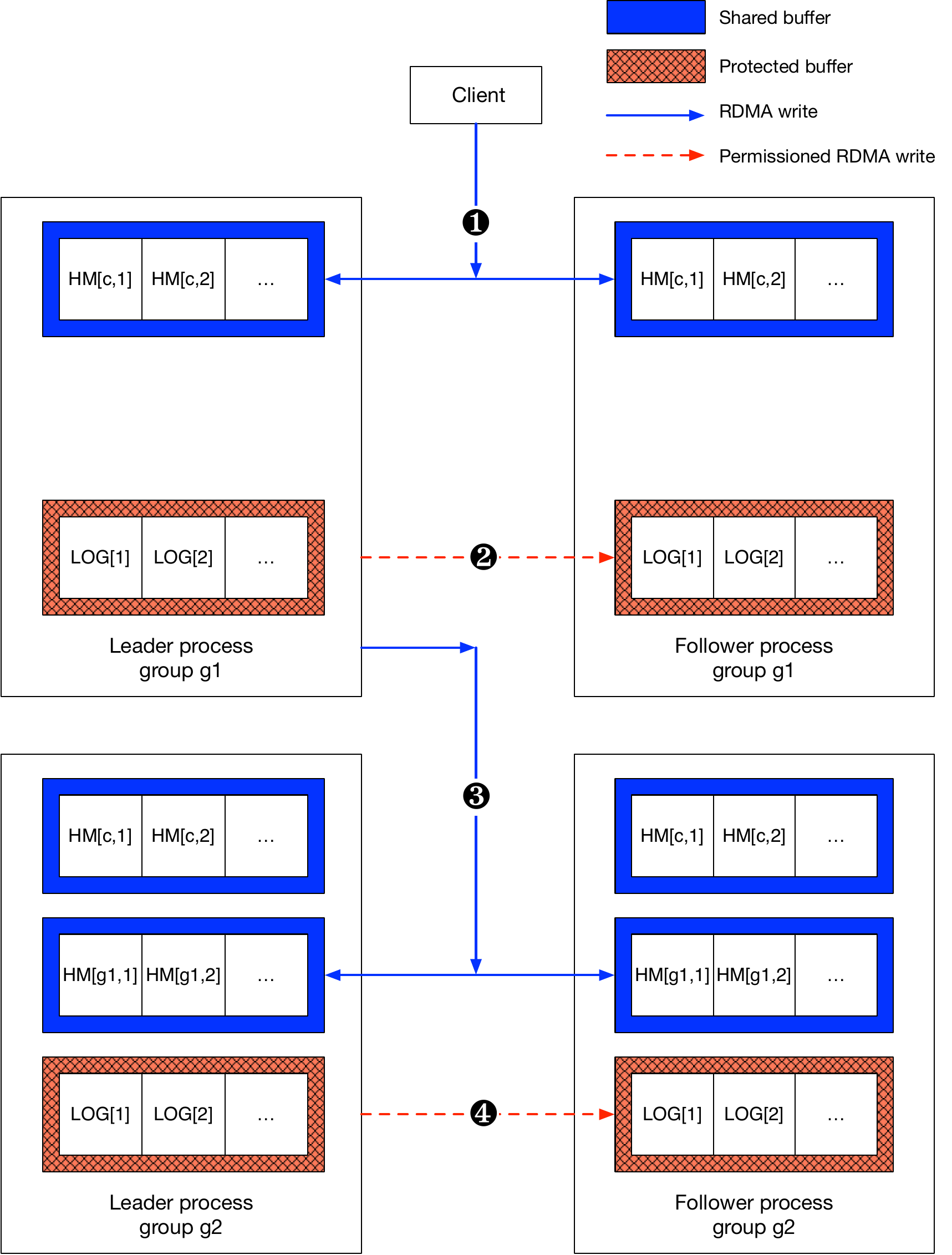}
  \captionsetup{font=small}
  \caption{A simple example with one client and groups $g_1$ and $g_2$, where $g_1$ is $g_2$'s parent, and two replicas per group, the leader and the follower. The client multicasts a message to both groups.}
  \label{fig:tree-example}
\end{figure}

In the absence of failures and with stable leaders, the algorithm is composed of the following steps (see Figure \ref{fig:tree-example}).
\begin{enumerate}
  \item[\circlednum{1}] A client atomically multicasts message $m$ by remotely writing $m$ in the buffer reserved for the client on all replicas of the \emph{lca} of $m$'s destinations.
  \item[\circlednum{2}] The leader of the \emph{lca} scans its buffers for new multicast messages, eventually finds $m$ and proposes an order for $m$ to the followers of its group by remotely updating their log.
  \item[\circlednum{3}] The leader delivers $m$ if the leader's group is a destination of the message (\libname is not genuine: messages ordered at a group may or may not have the group as a destination), and then remotely writes $m$ in the buffer of each of its children who have a destination within their reach.
  \item[\circlednum{4}] The leader of each group below the \emph{lca} scans its header buffers for new messages and runs Steps 2, 3, and 4 until all destination groups are reached.
\end{enumerate}

Intuitively, \libname orders messages consistently as follows.
If messages $m$ and $m'$ have the same \emph{lca}, then the \emph{lca} orders the messages and records their order in the group's log.
If $m$ and $m'$ do not have the same \emph{lca}, then there are two possibilities.
Case (i): $m$ and $m'$ do not have any destination in common, in which case their relative order is unimportant.
Case (ii): $m$ and $m'$ have a common destination. From the definition of \emph{lca}, this means that either $m'$'s \emph{lca} descends from $m$'s \emph{lca} or vice versa.
Without loss of generality, assume $m'$'s \emph{lca} descends from $m$'s \emph{lca}.
In this case, the relative order of $m$ and $m'$ will be defined by $m'$'s \emph{lca}, recorded in the log, and propagated down the tree if needed.
One important property of \libname is that descendant groups preserve the order established for any two messages by a group.
This happens because buffers shared by a parent group and a child group are read and written in order (i.e., FIFO order), hence the order of messages established by a parent group is preserved by its children.

\section{\libname detailed algorithm}
\label{sec:algodetails}

In this section, we detailed \libname.
Section~\ref{sec:datastr} introduces all shared data structures.
Section \ref{sec:client} describes how clients multicast messages, and Section \ref{sec:replicas} presents the main loop executed by the replicas.
Sections \ref{sec:perm}, \ref{sec:groupcu} and \ref{sec:treecu} explain how a replica replaces a failed or suspected leader.

%
%
%
%
%

\subsection{Shared data structures}
\label{sec:datastr}

Clients and replicas communicate using shared buffers \cite{FaRM}.
To simplify the description of \libname, we assume buffers are unbounded.
\libname's prototype introduces extensions to allow finite buffers, but these extensions do not fundamentally change the algorithms presented in this section.

\begin{itemize}
	\item $HM[0..|\mathcal{C}| + 1]$ is an array of buffers stored in the memory of replicas, with one buffer allocated for each client and one additional buffer for the parent of the group. Each buffer entry contains the message with its header, represented as a tuple $\langle c, id, dst, val \rangle$, where $c$ is the client that initiated the multicast, $id$ is a unique nonnull message identifier, $dst$ specifies the set of destination groups, and $val$ is the message payload.

	\item $LOG$ is a buffer stored in the memory of replicas. Each entry in the $LOG$ is a tuple $\langle tmp, source, slot_{out}, hm \rangle$, where $tmp$ is a system-wide unique timestamp needed to ensure consistency within a group (i.e., similar to a ballot in Paxos \cite{L98}), $source$ is the id of a client or the parent of the group, $slot_{out}$ is a vector that stores the position in the $HM$ of the children where the message is propagated down the tree, and $hm$ is a copy of the message the log entry refers to.

	\item $perm$ is an array with a timestamp for each remote replica, and it is used to allow a replica to ask for permissions to become leader by proposing a timestamp.

	\item $perm_{ack}$ is an array of pairs $\langle tmp, ack \rangle$ which will contain the result of the permission request for timestamp $tmp$, one for each remote replica.

\end{itemize}

\subsection{The client}
\label{sec:client}

Algorithm \ref{alg:client} shows how a client $c$ atomically multicasts a message $m$. First, $c$ computes the \emph{lca} of $m$'s destination set using the tree topology (i.e., the group from which all destinations can be reached by traversing the tree downwards).
Then, $c$ writes the message in the memory of the \emph{lca}'s replicas. Index $slot[x]$ points to the next free slot in each group $x$.

\begin{figure}[htbp]
	\centering
	\begin{algorithm}[H]
		\caption{\libname Client $c$}
		\label{alg:client}
		\begin{algorithmic}[1]

			\State \textbf{Initialization:}
			\Statex \hspace{1em} $\forall g \in \Gamma: slot[g] \gets 0$

			\vspace{2mm}

			\State \textbf{To multicast message $m = \langle id, dst, val \rangle$:}
			\State $lca \gets \texttt{lca}(\mathcal{T}, m.dst)$

			\ForAll{$p \in lca$}
			\State $\texttt{write}(p, HM[c, slot[lca]], \langle c, m.id, m.dst, m.val \rangle)$
			\EndFor
			\State $slot[lca] \gets slot[lca] + 1$

		\end{algorithmic}
	\end{algorithm}
\end{figure}

\subsection{The replicas}
\label{sec:replicas}

In addition to the shared data structures described in Section~\ref{sec:datastr}, replicas keep private variables:
$slot_{in}[x]$ points to the next slot to read from the memory of a client or the memory of the parent, and $slot_{out}[x]$ points to the next free slot to write in the memory of child $x$.
$FUO$, $STD$, $tmp$ and $leader$ are used to order messages in the group: $FUO$ (First Undecided Offset) points to the next log entry to decide, $STD$ (Slot To Deliver) points to the next log entry for which the delivery of the message must be handled, $tmp$ is a unique timestamp, and $leader$ is a local estimate of the current leader.
\libname orders messages within a group using a mechanism inspired by Paxos \cite{L98} and adapted to shared memory and permissions \cite{Mu}.

\algblockdefx{When}{EndWhen}[1]{\textbf{when} #1}{}
\algnotext{EndWhen}
\algblockdefx{Periodically}{EndPeriodically}[1]{\textbf{periodically do} #1}{}
\algnotext{EndPeriodically}

\begin{algorithm}
	\caption{\libname Main loop, process $s$ in group $g$}
	\label{alg:server}
	\begin{algorithmic}[1]

		\State \textbf{Initialization:}
		\Statex $\forall x \in \mathcal{C} \cup \lbrace \texttt{parent}(g) \rbrace: slot_{in}[x] \gets 0$
		\Statex $\forall x \in \texttt{children}(g): slot_{out}[x] \gets 0$
		\Statex $FUO, STD, tmp, leader \gets 0, 0, \langle r, 0 \rangle, \perp$
		\Statex $\forall r \in g: perm[r] \gets \langle 0, 0 \rangle$
		\Statex $\forall r \in g: perm_{ack}[r] \gets \langle \langle 0, 0 \rangle, \perp \rangle$

		\vspace{2mm}

		\When{$\texttt{leader(}g\texttt{)} = s \wedge leader \neq s$} \Comment{Step \ref{alg:server}.\ref{alg:server:1}}
		\State $res \gets \texttt{become\_leader}()$
		\State $FUO' \gets FUO$
		\State \textbf{if} $res = \texttt{OK}$ \textbf{then} $res \gets \texttt{group\_catchup}()$
		\If{$res = \texttt{OK}$}
		\State $\texttt{tree\_catchup}(FUO')$
		\State $leader \gets s$
		\EndIf
		\State \textbf{else} $leader \gets \perp$
		\EndWhen

		\vspace{2mm}

		\When{$leader = s\ \wedge$ \\ \hfill$\exists y\!\in\!\mathcal{C}\!\cup\!\lbrace \texttt{parent}(g) \rbrace\!:\!HM[y, slot_{in}[y]].id \neq 0$} \Comment{Step \ref{alg:server}.\ref{alg:server:2}} 

		\State $hm \gets HM[y, slot_{in}[y]]$
		\State $slot_{in}[y] \gets slot_{in}[y] + 1$

		\State $count \gets 0$ \Comment{Step \ref{alg:server}.\ref{alg:server:3}}
		\ForAll{$r \in g$}
		\State $l \gets \langle tmp, y, slot_{out}, hm \rangle$
		\State $res \gets \texttt{write}(r, LOG[FUO], l)$
		\If {$res = \texttt{OK}$}
		\State $count \gets count + 1$
		\EndIf
		\EndFor

		\If {$count > f$} \Comment{Step \ref{alg:server}.\ref{alg:server:4}}
		\State $STD \gets STD + 1$
		\If{$g \in hm.dst$}
		\State $\texttt{deliver}(hm.val)$\label{ln:deliver1}
		\EndIf

		\ForAll{$x \in \texttt{children}(g)$} \Comment{Step \ref{alg:server}.\ref{alg:server:5}}
		\If{$hm.dst \cap \texttt{reach}(x) \neq \emptyset$}
		\ForAll{$p \in x$}
		\State $\texttt{write}(p, HM[g, slot_{out}[x]], hm)$
		\EndFor
		\State $slot_{out}[x] \gets slot_{out}[x] + 1$
		\EndIf
		\EndFor

		\State $FUO \gets FUO + 1$

		\Else
		\State $leader \gets \perp$
		\EndIf
		\EndWhen

		\vspace{2mm}


		\When{$leader \neq s \wedge LOG[STD..STD+1].id \neq 0$} \Comment{Step \ref{alg:server}.\ref{alg:server:6}} 
		\State $l \gets LOG[STD]$
		\State $slot_{in}[l.source] \gets slot_{in}[l.source] + 1$
		\State $STD \gets STD + 1$
		\If{$g \in l.hm.dst$}
		\State $\texttt{deliver}(l.hm.val)$\label{ln:deliver2}
		\EndIf

		\EndWhen

		\vspace{2mm}

		\Periodically \Comment{Step \ref{alg:server}.\ref{alg:server:7}} 
		\ForAll{$r \in g$}
		\State $ts \gets perm[r]$ 
		\State $perm[r] \gets \langle 0, 0 \rangle$ 

		\If{$ts > tmp$}
		\State $tmp \gets ts$
		\State $\texttt{perm\_revoke}(leader)$
		\State $\texttt{perm\_grant}(r)$
		\State $\texttt{write}(r, perm_{ack}[s], \langle ts, \texttt{ACK} \rangle)$
		\State $leader \gets r$
		\EndIf

		\If{$ts < tmp$}
		\State $\texttt{write}(r, perm_{ack}[s], \langle ts, \texttt{NACK} \rangle)$
		\EndIf

		\EndFor
		\EndPeriodically

	\end{algorithmic}
\end{algorithm}

Algorithm \ref{alg:server} shows the behavior of a server $s$ in group $g$.
In the following, we comment on the main algorithm steps.

\begin{enumerate}
	\myitem{\ref{alg:server}} \label{alg:server:1} \emph{Becoming leader}: When replica $s$ queries $\texttt{leader(}g\texttt{)}$ and the output is its id, and $s$ is not the leader (i.e., $leader \neq s$), $s$ executes multiple steps. First, $s$ gathers permissions to read and write the log of the followers (i.e., $\texttt{become\_leader}()$), then $s$ ensures the replica logs are mutually consistent (i.e., $\texttt{group\_catchup}()$), and $s$ completes any operation the previous leader may have left unfinished because it failed or was demoted (i.e., $\texttt{tree\_catchup}()$).

	\myitem{\ref{alg:server}} \label{alg:server:2} \emph{Finding a message to be processed}: The leader checks the input buffer $HM[y, slot_{in}[y]]$ for each source $y$ to find a message ready to be processed and increments $slot_{in}[y]$ if it finds a message from $y$ .

	\myitem{\ref{alg:server}} \label{alg:server:3} \emph{Ordering messages}: The leader orders a message within its group by writing in the log of each replica in the group at slot $FUO$. By leveraging RDMA permissions, if the write succeeds in a quorum of replicas, the message is guaranteed to be ordered. If the leader fails to update a quorum, a new leader has taken over and revoked the previous leader's permissions.

	\myitem{\ref{alg:server}} \label{alg:server:4} \emph{Leader message delivery}: If the group of the leader is a destination, the leader delivers the message. In any case, the message has been checked for delivery, so $STD$ is incremented.

	\myitem{\ref{alg:server}} \label{alg:server:5} \emph{Forwarding to children}: For each child group that is a destination of the message or in the path to a destination, the message is written in the buffers of its replicas.

	\myitem{\ref{alg:server}} \label{alg:server:6} \emph{Followers message delivery}: Followers scan their log to check for the leader activity. When a follower finds messages in both slots $STD$ and $STD+1$, it can safely deliver the message in slot $STD$ (only if the group is a destination) \cite{Mu}.
	This holds because the leader writes slot $i+1$ if the message in slot $i$ has been ordered (i.e., successfully written in a quorum of replicas). $STD$ is incremented accordingly. Note that $STD \geq FUO$, since $FUO$ is only incremented when leader.

	\myitem{\ref{alg:server}} \label{alg:server:7} \emph{Checking permissions}:
	Replicas periodically check for permission requests issued by a leader candidate (described in Section \ref{sec:perm}).
	A replica grants permission and acknowledges the request if the candidate leader provides a timestamp greater than the largest timestamp the replica has received so far (as in phase 1 of Paxos \cite{L98}).
\end{enumerate}

\subsection{Gathering permissions}
\label{sec:perm}

When a replica attempts to become the leader, it first requests permission to update the log of the replicas (see Step \ref{alg:server}.\ref{alg:server:1} in Algorithm \ref{alg:server}).
Algorithm \ref{alg:perm} shows the steps executed by the candidate leader $s$ in group $g$.

\begin{figure}[htbp]
	\centering
	\begin{algorithm}[H]
		\caption{Procedure become\_leader(), process $s \in g$}
		\label{alg:perm}
		\begin{algorithmic}[1]

			\State $\texttt{increment}(tmp)$ \Comment{Step \ref{alg:perm}.\ref{alg:perm:1}}
			\ForAll{$r \in g$}
			\State $\texttt{write}(r, perm[s], tmp)$
			\EndFor

			\Repeat \Comment{Step \ref{alg:perm}.\ref{alg:perm:2}}
			\State $Q_{ACK}, Q_{NACK} \gets \emptyset, \emptyset$
			\ForAll{$r \in g$}
			\If{$perm_{ack}[r].ts = tmp$}
			\If{$perm_{ack}[r].ack = ACK$}
			\State $Q_{ACK} \gets Q_{ACK} \cup \lbrace r \rbrace$
			\Else
			\State $Q_{NACK} \gets Q_{NACK} \cup \lbrace r \rbrace$
			\EndIf
			\EndIf
			\EndFor
			\Until{$|Q_{ACK}| + |Q_{NACK}| > f$}

			\If{$|Q_{ACK}| \geq f + 1$}\Comment{Step \ref{alg:perm}.\ref{alg:perm:3}}
			\State \texttt{return} \texttt{OK}
			\EndIf
			\State \texttt{return} \texttt{NOK}

		\end{algorithmic}
	\end{algorithm}
\end{figure}


\begin{enumerate}
	\myitem{\ref{alg:perm}} \label{alg:perm:1} \emph{Proposing a timestamp}: Replica $s$ proposes a system-wide unique timestamp, greater than any timestamp it has proposed before, by writing the proposed timestamp in entry $perm[s]$ in the replicas.

	\myitem{\ref{alg:perm}} \label{alg:perm:2} \emph{Waiting for replies}: Replicas periodically scan $perm$ and if the proposed timestamp is greater than any timestamp the replica has read so far, grant permission and reply with an acknowledgement; otherwise with a negative acknowledgement (see Step \ref{alg:server}.\ref{alg:server:7}). Replica $s$ scans $perm_{ack}[s]$ waiting for a quorum of acknowledgements (positive or negative).

	\myitem{\ref{alg:perm}} \label{alg:perm:3} \emph{Checking the acknowledgements}: If $s$ collected a quorum of positive acknowledgements, it can assume to be the current leader.
\end{enumerate}

Even though multiple servers may attempt to become the leader simultaneously, at most one of them will have the correct permissions to write in a quorum of replicas to order a message.

\subsection{Group catchup}
\label{sec:groupcu}

Once a replica has obtained the necessary permissions to access the followers' logs, it ensures that the logs across replicas are consistent.
Algorithm \ref{alg:groupcu} shows the steps involved.

\begin{algorithm}
	\caption{Procedure group\_catchup(), process $s \in g$}
	\label{alg:groupcu}
	\begin{algorithmic}[1]

		\Loop 
		\State $count, L \gets 0, \emptyset$\Comment{Step \ref{alg:groupcu}.\ref{alg:groupcu:1}}
		\ForAll{$r \in g$}
		\State $res \gets \texttt{read}(r, LOG[FUO], l)$
		\State \textbf{if} $res \neq \texttt{OK}$ \textbf{then} $count \gets count + 1$
		\State \textbf{else} $L \gets L \cup \lbrace l \rbrace$
		\EndFor
		\State \textbf{if} $count > f$ \textbf{then} $\texttt{return NOK}$

		\vspace{2mm}
		\State $l_v \gets max_{tmp}(L)$ \Comment{Step \ref{alg:groupcu}.\ref{alg:groupcu:2}}
		\State \textbf{if} $l_v.tmp = 0$ \textbf{then} $\texttt{return OK}$

		\vspace{2mm}
		\State $count, l_v.tmp \gets 0, tmp$ \Comment{Step \ref{alg:groupcu}.\ref{alg:groupcu:3}}
		\ForAll{$r \in g$}
		\State $res \gets \texttt{write}(r, LOG[FUO], l_v)$
		\State \textbf{if} $res \neq \texttt{OK}$ \textbf{then} $count \gets count + 1$
		\State \textbf{if} $count > f$ \textbf{then} $\texttt{return NOK}$
		\EndFor

		\If{$FUO > STD$} \Comment{Step \ref{alg:groupcu}.\ref{alg:groupcu:4}}
		\State $STD \gets STD + 1$
		\If{$g \in l_v.hm.dst$}
		\State $\texttt{deliver}(l_v.hm.val)$\label{ln:deliver3}
		\EndIf
		\EndIf

		\vspace{2mm}
		\State $FUO \gets FUO + 1$ \Comment{Step \ref{alg:groupcu}.\ref{alg:groupcu:5}}
		\EndLoop

	\end{algorithmic}
\end{algorithm}

\begin{enumerate}
	\myitem{\ref{alg:groupcu}} \label{alg:groupcu:1} \emph{Reading remote logs}: The server tries to read the log from a quorum of remote replicas starting at entry $FUO$. If the replica reads from fewer than a quorum of replicas, it reverts to the follower logic.
	The replica fails to read from a quorum of followers when a new leader has taken over and revoked previous permissions.

	\myitem{\ref{alg:groupcu}} \label{alg:groupcu:2} \emph{Checking timestamps}: The replica picks the entry with the maximum timestamp among the entries read from a quorum. If no timestamp is found, the slot has not been decided yet, and the log is consistent with the replicas.

	\myitem{\ref{alg:groupcu}} \label{alg:groupcu:3} \emph{Replicating log entries}: If the maximum timestamp is not empty, the server replicates the entry with the largest timestamp in the log of the followers. As done before, if the remote write operations do not succeed in a quorum, a new leader has taken over.

	\myitem{\ref{alg:groupcu}} \label{alg:groupcu:4} \emph{Handle delivery}: Since the value at $FUO$ has been decided, it can be delivered if the current group is a destination. $STD$ is then incremented.

	\myitem{\ref{alg:groupcu}} \label{alg:groupcu:5} \emph{Moving on}: The replica concludes updating $FUO$, which points to the next entry available in the log.
\end{enumerate}

\subsection{Tree catchup}
\label{sec:treecu}

In addition to fixing log entries across replicas of its group, the leader must complete any operations left unfinished by the previous leader, such as updating the group's children.
Algorithm \ref{alg:treecu} details the steps involved.

\begin{algorithm}
	\caption{Procedure tree\_catchup($FUO'$), process $s \in g$}
	\label{alg:treecu}
	\begin{algorithmic}[1]

		\ForAll{$x \in \mathcal{C} \cup \lbrace \texttt{parent}(g) \rbrace$} \Comment{Step \ref{alg:treecu}.\ref{alg:treecu:1}}
		\State $last \gets max(\{i : LOG[i].source = x\})$
		\State $slot_{in}[x] \gets i + 1 : HM[i].id = LOG[last].id$
		\EndFor

		\vspace{2mm}

		\ForAll{$last \in [FUO' .. FUO - 1]$} \hfill \Comment{Step \ref{alg:treecu}.\ref{alg:treecu:2}}
		\State $l_{last} \gets LOG[last]$
		\ForAll{$x \in \texttt{children}(g)$}
		\State $slot_{out}[x] \gets l_{last}.slot_{out}[x]$




		\If{$l_{last}.hm.dst \cap \texttt{reach}(x) \neq \emptyset$} \hfill \Comment{Step \ref{alg:treecu}.\ref{alg:treecu:3}}
		\ForAll{$p \in x$}
		\State $\texttt{write}(p, HM[g, slot_{out}[x]], l_{last}.hm)$
		\EndFor
		\State $slot_{out}[x] \gets slot_{out}[x] + 1$
		\EndIf
		\EndFor
		\EndFor

	\end{algorithmic}
\end{algorithm}

\begin{enumerate}
	\myitem{\ref{alg:treecu}} \label{alg:treecu:1} \emph{Updating input pointers}: For each input $x$ (i.e., parent and clients), $slot_{in}[x]$ is updated by finding the message id in the last log slot associated with $x$. The slot in $HM[x]$ with the same message id is the last consumed one, so we move to the next.

	\myitem{\ref{alg:treecu}} \label{alg:treecu:2} \emph{Multicast logic}: Since the old leader may have been faulty, there is no guarantee that the values in the log have been replicated to the children. The new leader executes the multicast logic again for each log slot from the value of $FUO$ before the failover. For each child group $x$, the server updates $slot_{out}[x]$ with the value stored in the log.

	\myitem{\ref{alg:treecu}} \label{alg:treecu:3} \emph{Re-replicating on children}: The server rewrites the entry and increments $slot_{out}[x]$ in group $x$, if the group is a destination or in the path to a destination.
\end{enumerate}

\section{\libname Correctness}
\label{sec:proof}

\newtheorem{theorem}{Theorem}
\newtheorem{lemma}{Lemma}
\newtheorem{proposition}{Proposition}

This section argues that \libname ensures all atomic multicast properties (Section \ref{sec:amcast}).
We initially prove properties of messages multicast to a single group, and then show how \libname builds on these properties to implement atomic multicast.

We refer to the event of multicasting a message $m$ to a group $x$ as $x$-broadcast of $m$, and the event of delivering $m$ as $x$-deliver of $m$.
The event of $x$-broadcasting $m$ happens when a client multicasts $m$ and $x=lca(m)$ (Algorithm \ref{alg:client}) or when a process $p$ assesses that $x$ is $m$'s destination or in the path to $m$'s destinations (Algorithm \ref{alg:server}).

Since follower processes in a group deliver a message after they learn the existence of a subsequent message in their log, we assume executions in which an infinite number of messages are multicast to each group.
Circumventing this assumption is straightforward but would add complexity to the algorithms.

\vspace{2mm}
\begin{lemma}
  In every group $x$, eventually some correct process $p$ becomes a stable leader (i.e., $p$ is elected leader and never demoted).
  \label{lemma:aux1}
\end{lemma}
\vspace{2mm}
\noindent
{\sc Proof:}
From the definition of the $leader()$ oracle (Section \ref{sec:system-model}), there is a time $t_1$ after which all correct processes output the same leader $p$ in $x$.
Moreover, from the definition of correct processes, there is a time $t_2$ after which only correct processes execute.
Thus, after time $t = max(t_1,t_2)$ all processes in $x$ in execution output $p$.
\hfill$\Box$

\vspace{2mm}
\begin{lemma}
  In every group $x$, a stable leader eventually permanently acquires write permission in the log of all correct followers.
  \label{lemma:aux2}
\end{lemma}
\vspace{2mm}
\noindent
{\sc Proof:}
Assume for a contradiction that stable leader $p$ never permanently acquires write permission in the log of all correct followers.
To acquire write permissions, $p$ must execute procedure become\_leader() (Algorithm \ref{alg:perm}) successfully (i.e., return OK).
Thus, the procedure always returns NOK and $p$ never sets variable $leader$ to $p$ (Step \ref{alg:server}.\ref{alg:server:1}).
From Lemma~\ref{lemma:aux1}, there is some time after which $p$ constantly re-executes Step \ref{alg:server}.\ref{alg:server:1} in Algorithm \ref{alg:server} and no other process executes the step, but $p$ never becomes the leader.
Each time $p$ calls procedure become\_leader() (Algorithm \ref{alg:perm}), $p$ increments its timestamp variable $tmp$, and since no other process in $x$ executes become\_leader(), eventually $tmp$ has a value greater than any other previously proposed timestamp.
When $p$ executes become\_leader() with such a large timestamp value, when executing Step \ref{alg:server}.\ref{alg:server:7}, a follower revokes any existing permissions and sets the permission to $p$, a contradiction.
\hfill$\Box$

\vspace{2mm}
\begin{proposition}
  \textit{Group validity}:~if a correct process $p$ $x$-broadcasts a message $m$, all correct processes $q \in x$ eventually $x$-deliver $m$.
  \label{lemma:Tg1}
\end{proposition}
\vspace{2mm}
\noindent
{\sc Proof:}
Process $p$ $x$-broadcasts $m$ to $x$ in two cases:
(i) $p$ is a client process and $x$ is the $lca$ of $m.dst$ (Algorithm~\ref{alg:client}), and
(ii) $p$ is a process in the group that is the parent of $x$ and $x$ is a destination of $m$ or $x$ is in the path to a destination of $m$ (Algorithm~\ref{alg:server}, Step \ref{alg:server}.\ref{alg:server:5}).
In both cases, since $p$ is correct, it writes $m$ in the $HM$ data structure of all correct processes in $x$.

There are three cases in which processes deliver a message:
(a) when a leader executes line \ref{ln:deliver1} in Algorithm~\ref{alg:server},
(b) when a follower executes line \ref{ln:deliver2} in Algorithm~\ref{alg:server}, and
(c) when a new leader executes line \ref{ln:deliver3} in Algorithm \ref{alg:groupcu}.
The proof continues by contradiction.

Case (a). Assume no leader $q$ ever delivers $m$.
Since $m$ has been written in $q$'s $HM$ data structure, by Step \ref{alg:server}.\ref{alg:server:2}, $q$ will eventually return $m$.
From Step \ref{alg:server}.\ref{alg:server:3} in Algorithm~\ref{alg:server}, $q$ never manages to write $m$ in an entry in the log of a quorum of processes in $x$.
Since there is a quorum of correct processes, it must be that when executing a remote write in Step \ref{alg:server}.\ref{alg:server:3} in the memory of process $r$, $q$ does not have $r$'s permission.
From Lemma \ref{lemma:aux2}, there is a stable leader $r$ that will eventually have permanent permissions, a contradiction.

Case (b). Assume no follower $q$ delivers $m$.
Then, either:
(b.1) $m$ is not included in its log,
(b.2) $m$ is included in its log, but there is no message in entry $FUO+1$, or
(b.3) $q$ becomes the leader before it executes Step \ref{alg:server}.\ref{alg:server:3}.
From Case (a), when $r$ becomes a stable leader, it will copy all entry logs to the followers' log.
From the assumption that there is an unbounded number of messages multicast to $x$, eventually there will be a message in $q$'s log entry $FUO+1$.
Therefore, it must be that $q$ becomes the leader before it delivers $m$.
This is covered in case (c) discussed next.

Case (c). When $q$ executes Algorithm \ref{alg:groupcu}, either (c.1) $m$ has already been written in a quorum of processes by a previous leader or (c.2) not.
In case (c.1), $q$ will deliver $m$ upon executing Step \ref{alg:groupcu}.\ref{alg:groupcu:3}.
In case (c.2), $q$ will deliver $m$ as the leader as in case (a) or as a follower as in case (b), a contradiction that concludes the proof.
\hfill$\Box$

\vspace{2mm}
\begin{proposition}
  \textit{Group integrity}:~for any process $p$ and any message $m$, $p$ $x$-delivers $m$ at most once, and only if $p \in x$ and $m$ was $x$-broadcast.
  \label{lemma:Tg2}
\end{proposition}
\vspace{2mm}
\noindent
{\sc Proof:}
A process $p$ can $x$-deliver $m$ in three situations:
(i)~as leader in Step \ref{alg:server}.\ref{alg:server:4},
(ii)~as follower in Step \ref{alg:server}.\ref{alg:server:6}, and
(iii)~while becoming leader in Step \ref{alg:groupcu}.\ref{alg:groupcu:3}.
From Algorithm \ref{alg:server}, processes $x$-deliver messages stored in the log at the slot indexed by $STD$ and increment $STD$ after $x$-delivering a message.
From Algorithm \ref{alg:groupcu}, a process never $x$-delivers a message placed in a slot smaller than $STD$.
Thus, if $m$ is stored in at most one entry in the log, it will not be $x$-delivered twice.

We now show that $m$ is not stored in two entries in the log.
From Algorithm~\ref{alg:server}, before $x$-delivering $m$, leader $p$ increments $slot_{in}[x]$ and creates an entry for $m$ in its local and remote logs.
By incrementing $slot_{in}[x]$, $p$ will not include $m$ in the log again in Step \ref{alg:server}.\ref{alg:server:4}.
When a follower $q$ becomes the leader, it first executes Algorithm \ref{alg:groupcu} to ensure processes have consistent logs.
And the Algorithm \ref{alg:treecu}, to guarantee that if $m$ is already in $q$'s log, say in entry $k$, then is set to a value $slot_{in}[x]$ greater than $k$ (Step \ref{alg:treecu}.\ref{alg:treecu:1}).

%
%

Processes only $x$-deliver messages in $HM$, and since a message $m$ is only written in a process's $HM$ data structure by a client when the client multicasts $m$ (Algorithm \ref{alg:client}) and by the leader of a group if the child is a destination of $m$ or in a path to $m$'s destination (Step \ref{alg:server}.\ref{alg:server:5} in Algorithm~\ref{alg:server}), we conclude that every $x$-delivered message has been $x$-broadcast.
\hfill$\Box$

\vspace{2mm}
\begin{proposition}
  \textit{Group agreement}:~if a process $p$ $x$-delivers a message $m$, all correct processes $q$ in group $x$ $x$-deliver $m$.
  \label{lemma:Tg3}
\end{proposition}
\vspace{2mm}
\noindent
{\sc Proof:}
Assume $p$ $x$-delivers $m$, either as (i)~leader or (ii)~follower (lines \ref{ln:deliver1} or \ref{ln:deliver2} in Algorithm~\ref{alg:server}, respectively), or (iii)~when becoming leader (line \ref{ln:deliver3} in Algorithm \ref{alg:groupcu}).

Case (i). If $p$ $x$-delivers $m$ as leader, then $p$ has successfully written $m$ in the $k$-th entry in the log (i.e., $FUO = k$) of a quorum $Q$ of processes (Steps \ref{alg:server}.\ref{alg:server:3} and \ref{alg:server}.\ref{alg:server:4}).
Assume for a contradiction that $q$ never $x$-delivers $m$.
If $q$ is a correct process in $Q$, then it will eventually find an entry in its log that succeeds $k$ and $x$-deliver $m$ (Step \ref{alg:server}.\ref{alg:server:6}), a contradiction.
Thus, assume that $q$ is a correct process not in $Q$.
For this to happen, $p$ must have failed or been demoted before it updated $q$'s log.
Let $r$ be the next leader that successfully sets $FUO = k+1$ in group\_catchup() (Algorithm \ref{alg:groupcu}).
Process $r$ exists from Lemmas \ref{lemma:aux1} and \ref{lemma:aux2}.
Since $p$ has updated a quorum of processes, $r$ reads at least one log such that the $k$-th entry contains $m$.
From Step \ref{alg:groupcu}.\ref{alg:groupcu:3}, $r$ updates the $k$-th log entry of all correct processes with $m$ and $x$-delivers $m$.
We conclude that $q$ must be a follower and, from the discussion above, $q$ has $m$ in the $k$-th entry of its log.
When a subsequent message is included in $q$'s log, from Step \ref{alg:server}.\ref{alg:server:6}, $q$ will $x$-deliver $m$, a contradiction.

Case (ii). Assume $p$ $x$-delivers $m$ as a follower.
From Step \ref{alg:server}.\ref{alg:server:6}, $m$ must be in some entry, say $k$, in $p$'s log, and there must be a message $m'$ in the $(k+1)$-th entry in $p$'s log.
Therefore, by Step \ref{alg:server}.\ref{alg:server:3} and Lemma \ref{lemma:aux2}, there is a leader $r$ that updates the $k$-th entry of all correct processes $q$ with $m$.
From Step \ref{alg:server}.\ref{alg:server:6}, $q$ will eventually deliver $m$.

Case (iii). Consider $p$ $x$-delivers $m$ when executing Algorithm \ref{alg:groupcu}.
Then, from Step \ref{alg:groupcu}.\ref{alg:groupcu:3}, $p$ has updated the log of a quorum of processes with $m$, and in Case (i), $q$ $x$-delivers $m$.
\hfill$\Box$

\vspace{2mm}
\begin{proposition}
  \textit{Group total order}:~no two processes $x$-deliver any two messages in different orders.
  \label{lemma:Tg4}
\end{proposition}
\vspace{2mm}
\noindent
{\sc Proof:}
Total order follows immediately from the fact that
(i) each $x$-delivered message is stored in the same position (i.e., $FUO$) in the log of processes and
(ii) processes $x$-deliver messages in order of their position in the log.
\hfill$\Box$

%
%

\vspace{2mm}
\begin{lemma}
  For any message $m$ multicast to multiple groups, let group $x_0$ be the lowest common ancestor of $m.dst$.
  For all $x_d \in m.dst$, if correct process $p$ in $x_0$ $x_0$-delivers $m$, then all correct processes in the path $x_1, ..., x_d$, $x_k$-deliver $m$, where $1 \leq k \leq d$.
  \label{lemma:T1}
\end{lemma}

\vspace{2mm}
{\sc Proof:}
By induction.
(Base step.)
Since correct $p$ $x_0$-delivers $m$, $x_1$ is a child of $x_0$, and $\mathit{reach}(x_1) \cap m.dst \neq \emptyset$, $p$ $x_1$-broadcasts $m$.
The claim follows from group validity of $x_1$.
(Inductive step.)
Assume each correct process $r$ in $x_k$ $x_k$-delivers $m$.
From Algorithm \ref{alg:server}, and the fact that $x_{k+1}$ is a child of $x_{k}$ and $\mathit{reach}(x_{k+1}) \cap m.dst \neq \emptyset$, $r$ $x_{k+1}$-broadcasts $m$.
From the group validity of $x_k$, every correct process in $x_{k+1}$ $x_{k+1}$-delivers $m$.
\hfill$\Box$
\vspace{2mm}

\begin{lemma}
  For any multicast message $m$, let group $x_0$ be the lowest common ancestor of $m.dst$.
  For all $x_d \in m.dst$, if correct process $p$ in $x_d$ $x_d$-delivers $m$, then all correct processes in the path $x_0, ..., x_d$, $x_k$-deliver $m$, where $0 \leq k \leq d$.
  \label{lemma:T2}
\end{lemma}
\vspace{2mm}
{\sc Proof:} By backwards induction.
(\emph{Base step.}) The case for $k=d$ follows directly from group agreement in group $x_d$.
(\emph{Inductive step.})
Assume that every correct process $r \in x_k$ $x_k$-delivers $m$.
We show that correct processes in $x_{k-1}$ $x_{k-1}$-deliver $m$.
From group integrity in $x_k$, one correct process $s$ in $x_{k-1}$ $x_k$-broadcasts $m$.
Therefore, $s$ $x_{k-1}$-delivers $m$, and from group agreement in $x_{k-1}$ all correct processes $x_{k-1}$-deliver $m$.
\hfill$\Box$
\vspace{2mm}

\begin{proposition}
  (\textit{Validity}) If a correct client process $p$ multicasts a message $m$, then eventually all correct processes $q \in g$, where $g \in m.\mathit{dst}$, deliver $m$.
  \label{prop:P1}
\end{proposition}
\vspace{2mm}
{\sc Proof:}
Let group $x_0$ be the lowest common ancestor of $m.dst$ and $x_d$ a group in $m.dst$.
From Algorithm \ref{alg:client}, $p$ $x_0$-broadcasts $m$ and from group validity, all correct processes in $x_0$ $x_0$-deliver $m$.
From Lemma~\ref{lemma:T1}, all correct processes in $x_d$, $x_d$-deliver $m$.
Hence, every correct process in $x_d$ a-delivers $m$.
\hfill$\Box$
\vspace{2mm}

\begin{lemma}
  If process $p$ $x$-delivers a message $m$, then eventually a correct process $q$ $x$-delivers $m$.
  \label{lemma:T5}
\end{lemma}
\vspace{2mm}
{\sc Proof:} $p$ can deliver $m$ (i) as leader at Step \ref{alg:server}.\ref{alg:server:4}, (ii) as follower at Step \ref{alg:server}.\ref{alg:server:6}, or (iii) while becoming leader at Step \ref{alg:groupcu}.\ref{alg:groupcu:4}. In all cases, since writes to the log are permissioned, $m$ was already replicated in a quorum of processes: at Step \ref{alg:server}.\ref{alg:server:3} in case (i), at Step \ref{alg:groupcu}.\ref{alg:groupcu:3} in case (iii), and because a leader does not move to the next $FUO$ before replicating the previous one in case (ii). From Lemma~\ref{lemma:aux1} and Lemma~\ref{lemma:aux2}, some correct process $q$ eventually becomes a stable leader, finds $m$ in its log and $x$-delivers $m$.
\hfill$\Box$
\vspace{2mm}

\begin{proposition}
  (\textit{Uniform agreement})~If a process $p$ in group $x_d$ delivers a message $m$, then eventually all correct processes $q \in g$, where $g \in m.\mathit{dst}$, deliver $m$.
  \label{prop:P2}
\end{proposition}
\vspace{2mm}
{\sc Proof:}
From Lemma~\ref{lemma:T5} and Lemma~\ref{lemma:T2}, all correct processes in $x_0$, $x_0$-deliver $m$.
Thus, from Lemma~\ref{lemma:T1}, all $x_d \in m.dst$ $x_d$-deliver $m$.
It follows from Algorithm \ref{alg:server} that all $q \in x_d$ a-deliver $m$.
\hfill$\Box$
\vspace{2mm}

\begin{proposition}
  (\textit{Uniform integrity})~For any process $p$ and any message $m$, $p$ delivers $m$ at most once, and only if $p \in g$, $g \in m.\mathit{dst}$, and $m$ was previously multicast.
  \label{prop:P3}
\end{proposition}
\vspace{2mm}
{\sc Proof:}
If $m$ is multicast to a single group, then the proposition follows immediately from group integrity (Proposition \ref{lemma:Tg2}).

Let $x$ and $y$ be two groups in $m$'s destination or the path to $m$'s destination such that $x$ is $y$'s direct descendant.
We show processes in $y$ $x$-broadcast $m$ only once.
There are two cases to consider:

Case (i). Assume $p \in y$ $x$-broadcasts $m$ at Step \ref{alg:server}.\ref{alg:server:5}.
This is triggered because $p$ found $m$ in entry $slot_{in}[z]$ in $HM[z]$, for some group $z$.
Since $p$ increments $slot_{in}[z]$, $p$ will not $x$-broadcast $m$ again at Step \ref{alg:server}.\ref{alg:server:5}.
From Step \ref{alg:treecu}.\ref{alg:treecu:1}, any other process $q$ that becomes the leader in $y$ will update $slot_{in}[z]$ similarly to $p$, so $q$ will not $x$-broadcast $m$ again at Step \ref{alg:server}.\ref{alg:server:5}.

Case (ii). Consider now that $p$ $x$-broadcasts $m$ at Step \ref{alg:treecu}.\ref{alg:treecu:3}.
In this case, in case (i), $p$ will place $m$ in the same entry in $HM[x]$ as the previous leader.
\hfill$\Box$
\vspace{2mm}

We define $\mathit{subtree}(m)$ as the subtree of the overlay tree $\mathcal{T}$ rooted at $lca(m)$ and descending to all destinations of $m$.

\begin{lemma}
  If $m$ and $m'$ are two messages multicast to one or more destination groups in common, then
  $lca(m) \in \mathit{subtree}(m')$ or $lca(m') \in \mathit{subtree}(m)$.
  \label{lemma:T3}
\end{lemma}
\vspace{2mm}
{\sc Proof:}
Assume group $x$ is a common destination in $m$ and $m'$ (i.e., $x \in m.dst \cap m'.dst$).
Let $path(x)$ be the sequence of groups in the overlay tree $\mathcal{T}$ from the root until $x$.
From Algorithm \ref{alg:server}, in order to reach $x$, $lca(m)$ (resp., $lca(m')$) must be a group in $path(x)$.
Without loss of generality, assume that $lca(m)$ is higher than $lca(m')$ or at the same height as $lca(m')$.
Then, $lca(m') \in \mathit{subtree}(m)$, which concludes the lemma.
\hfill$\Box$
\vspace{2mm}

\begin{lemma}
  If a correct process in group $x_0$ $x_0$-delivers $m$ before $m'$, then for every ancestor group $x_d$ of $x_0$, where $x_d \in m.dst \cap m'.dst$, every correct process in $x_d$ $x_d$-delivers $m$ before $m'$.
  \label{lemma:T4}
\end{lemma}
\vspace{2mm}
{\sc Proof:}
By induction on the path $x_d, ..., x_{k+1}, x_k, ..., x_0$.
(\emph{Base step.})
Trivially from the properties of broadcast in group $x_0$.
(\emph{Inductive step.})
Let $p \in x_k$ $x_k$-deliver $m$ before $m'$.
Since $x_d \in m.dst \cap m'.dst$, and from Lemma~\ref{lemma:T5}, some correct process $q \in x_{k+1}$ $x_k$-broadcasts $m$ before $m'$ and $m$ precedes $m'$ in the log of processes in $x_{k+1}$.
It follows that every correct process $q \in x_{k+1}$ $x_{k+1}$-delivers $m$ before $m'$.
\hfill$\Box$

\vspace{2mm}
\begin{proposition}
  (\textit{Prefix order})~For any two messages $m$ and $m'$ and any two correct processes $p$ and $q$ such that $p \in g$, $q \in h$ and $\{ g, h \} \subseteq m.\mathit{dst} \cap m'.\mathit{dst}$, if $p$ delivers $m$ and $q$ delivers $m'$, then either $p$ delivers $m'$ before $m$ or $q$ delivers $m$ before $m'$.
  \label{prop:P4}
\end{proposition}
\vspace{2mm}
{\sc Proof:}
The proposition holds trivially if $p$ and $q$ are in the same group, so assume that $g \neq h$.
From Lemma~\ref{lemma:T3}, and without loss of generality, assume that $lca(m') \in \mathit{subtree}(m)$.
Thus, $lca(m')$ will order $m$ and $m'$.
From Lemma~\ref{lemma:T4}, both $p$ and $q$ a-deliver $m$ and $m'$ in the same order as $lca(m')$.
\hfill$\Box$

\vspace{2mm}
\begin{proposition}
  (\textit{Acyclic order})~The relation $<$ is acyclic.
  \label{prop:P5}
\end{proposition}
\vspace{2mm}
{\sc Proof (sketch):}
For a contradiction, assume there is an execution of \libname that results in a cycle $m_0 < ... < m_d < m_0$.
Since all correct processes in the same group deliver messages in the same order, the cycle must involve messages multicast to multiple groups.
Let $x$ be the highest lowest common ancestor of all messages in the cycle.
We define $\mathit{subtree}(x,1)$, $\mathit{subtree}(x, 2), ...$ as the subtrees of group $x$ in $\mathcal{T}$.
Since the cycle involves groups in the subtree of $x$, there must exist messages $m$ and $m'$ such that
(a)~$m$ is delivered before $m'$ in groups in $\mathit{subtree(x, i)}$ and
(b)~$m'$ is delivered before $m$ in groups in $\mathit{subtree(x, j)}, i \neq j$.
From Lemma~\ref{lemma:T4}, item (a) implies that processes in $x$ $x$-deliver $m$ and then $m'$, and item (b) implies that processes in $x$ $x$-deliver $m'$ and then $m$, a contradiction.
\hfill$\Box$

\section{Implementation}
\label{sec:implementation}

\libname was implemented in C++ using \texttt{g++} version 11.4.0 and \texttt{cmake} version 3.22.1. \libname's prototype uses a custom library that enables processes to share queue pairs (via TCP), facilitating the sharing of local variables, remote variable access (read/write), and dynamic adjustment of associated permissions. This library is built on the \texttt{rdma/rdma\_cma.h} Linux library.

To have full and immediate visibility of data written by remote processes, we label all shared variables as \texttt{volatile}, thus preventing the compiler from optimizing or caching local reads and writes. Small remote writes are \emph{inlined}, meaning the data is kept within the network packet to optimize the DMA accesses to the local memory by the RNIC. The inline limit varies among different machines; in our case, it is approximately 1 Kilobyte. 

One of the main challenges of shared memory implementations is correctly handling concurrency between remote writes and local reads to prevent race conditions. This is typically achieved by adding a canary value or a checksum to the written memory region \cite{APUS, Mu}. Since computing a checksum is expensive compared to RDMA operations performance, we used a canary approach that leverages left-to-right semantics of read and write operations together with the fact that operations of the size of a cache line are atomic when memory is aligned. Accordingly, the last elements in the entries we write to remote buffers are always smaller than a cache line and act as our canaries. Note that left-to-right semantics of read and write operations is not granted by default with RDMA, although most NICs allow the user to pin memory regions to NUMA cores to ensure this.

\section{Performance evaluation}
\label{sec:perfeval}

%

\subsection{Preliminaries}

To evaluate the performance of \libname and compare it with current state-of-the-art we leveraged the CloudLab \cite{Duplyakin+:ATC19} public test-bed, allocating a set of R320 nodes running Ubuntu Linux 22.04 LTS, equipped with one 8-core Xeon E5-2450 processor, 16 GB of main memory, and a Mellanox FDR CX3 NIC. The network has a nominal Infiniband write latency of $1.31 \mu s$ (from \verb|ib_write_lat|) and bandwidth of $50 Gbps$ (from \verb|ib_write_bw|).

\begin{figure}[htbp]
  \centering
  \begin{tabular}{c@{\hskip 0.1em}c@{\hskip 2em}c}
    \raisebox{-.5\height}{\includegraphics[width=0.18\textwidth]{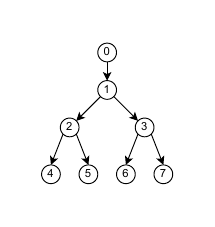}}    &
    \raisebox{-.5\height}{\includegraphics[width=0.18\textwidth]{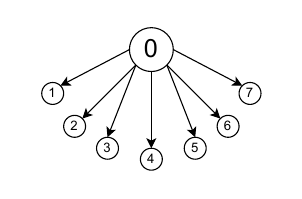}} &
    \raisebox{-.5\height}{\includegraphics[width=0.03\textwidth]{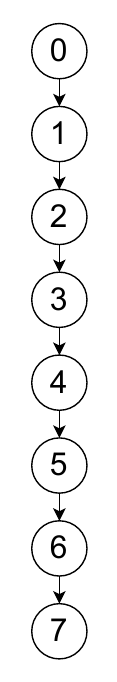}}     \\
  \end{tabular}
  \caption{\libname overlays: base, breadth, and depth (left to right).}
  \label{fig:trees}
\end{figure}

In our evaluation, we consider three overlays for \libname: base, breadth, and depth (see Figure~\ref{fig:trees}) and compare them with RamCast \cite{le2021ramcast}, the state-of-the-art shared-memory genuine atomic multicast protocol.
Unless otherwise noted, each group is composed of three processes: the leader and two followers.
For a fair comparison, we reimplemented RamCast using the same language (C++), custom library (see previous section), and optimizations as \libname's. The performance results from our RamCast implementation are comparable to the original paper results \cite{le2021ramcast}, with marginal differences consistent with the differences between our deployment and the original one (nodes and network).

\subsection{Impact of tree design under a random workload}
\label{sec:perf:design}

\begin{figure}[htbp]
  \centering
  \begin{tabular}{ccc}
    \includegraphics[width=\mygraphsize\textwidth]{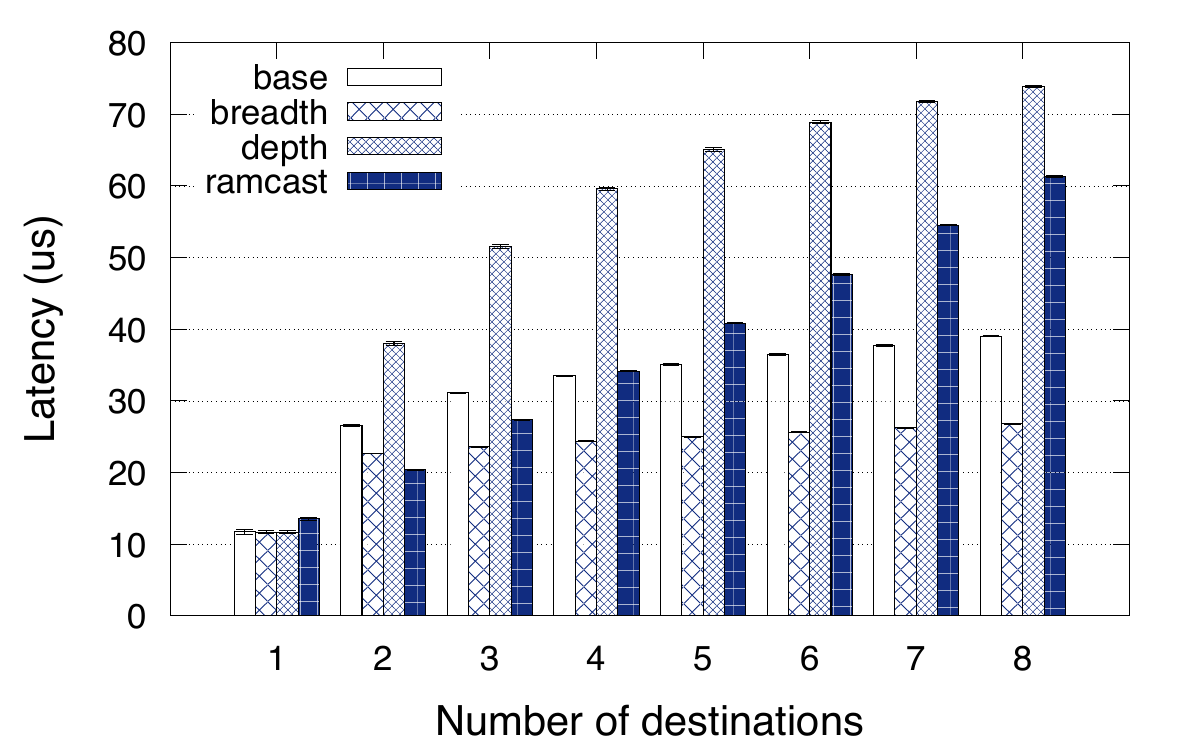} &
    \includegraphics[width=\mygraphsize\textwidth]{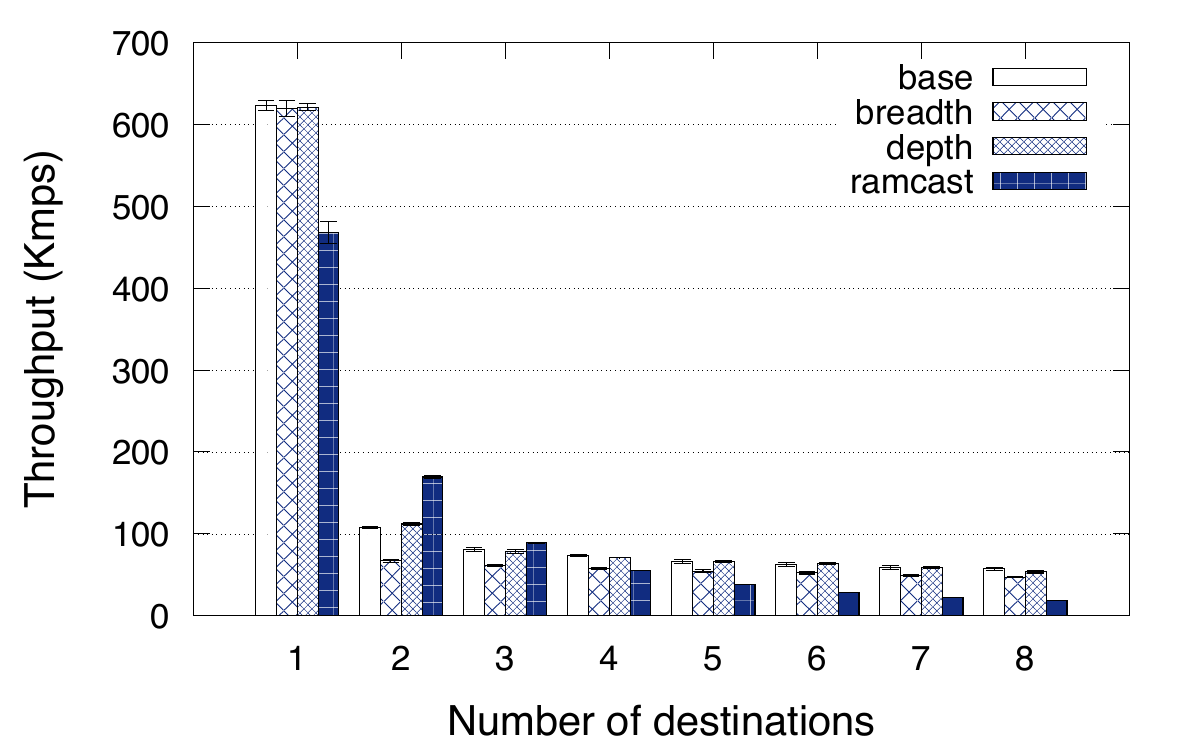} &
    \includegraphics[width=\mygraphsize\textwidth]{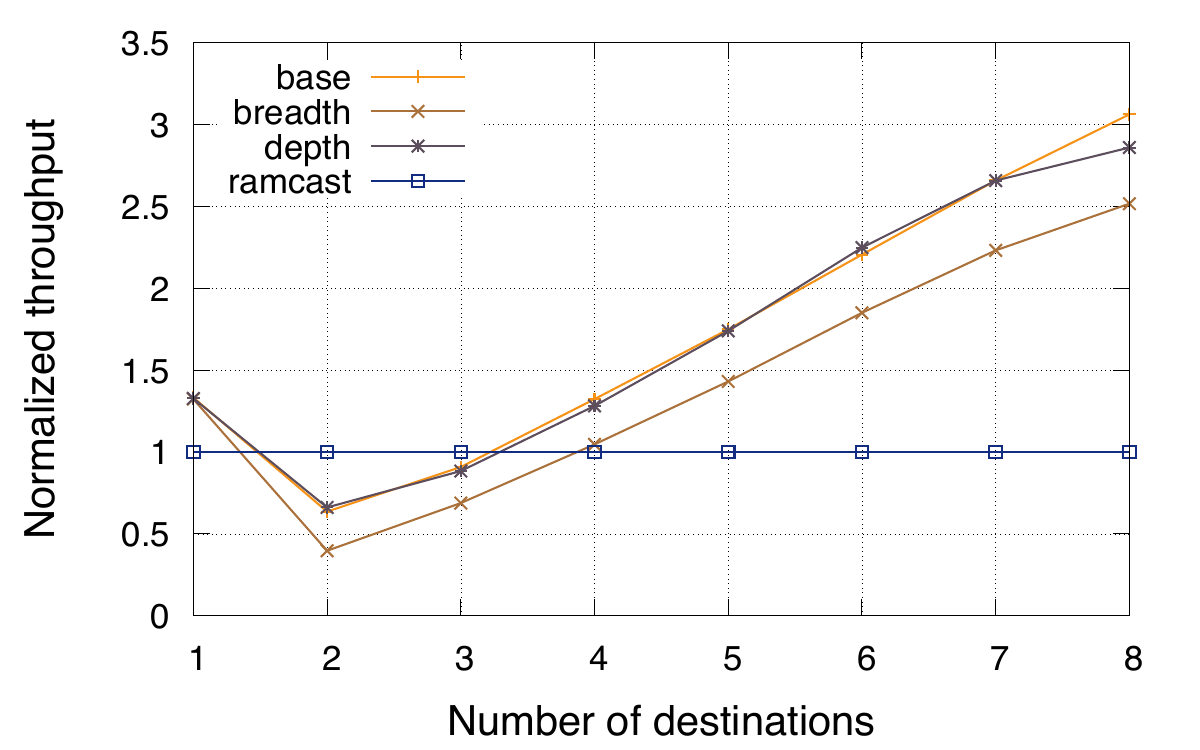} \\
  \end{tabular}
  \caption{Latency (top), throughput (center), and normalized throughput (bottom) of \libname and RamCast under a randomly uniform workload with 8 groups (3 replicas each) and 64B payload. Latency measured with one client; throughput at system saturation.}
  \label{fig:perf:design}
\end{figure}

Figure \ref{fig:perf:design} compares \libname with RamCast considering a workload where messages are multicast to one or more destinations chosen randomly (following a uniform distribution) among the eight possible destinations we consider.
Results show that the \emph{breadth} tree can order messages almost in constant time, excluding the faster case of a single destination, regardless of the number of destinations. This is a result of optimizing write operations to children (see Section \ref{sec:design:motiv}): once the \emph{lca} ordered the message, all children can process it in parallel. \emph{Base} and \emph{depth} trees have a higher latency due to the need to perform additional hops to reach all destinations.
In terms of throughput, the \emph{depth} tree achieves best performance (closely followed by the \emph{base} tree) due to its ability to parallelize multicast message ordering across multiple \emph{lcas} distributed throughout the tree.
RamCast's genuineness gives it an advantage over \libname with few destinations. However, choosing the ``right'' tree for the case, \libname achieves lower latency for more than two destinations and higher throughput for more than three.

\subsection{Skewed, ``genuine'' workload}
\label{sec:perf:skew}

\begin{figure}[htbp]
  \centering
  \begin{tabular}{ccc}
    \includegraphics[width=\mygraphsize\textwidth]{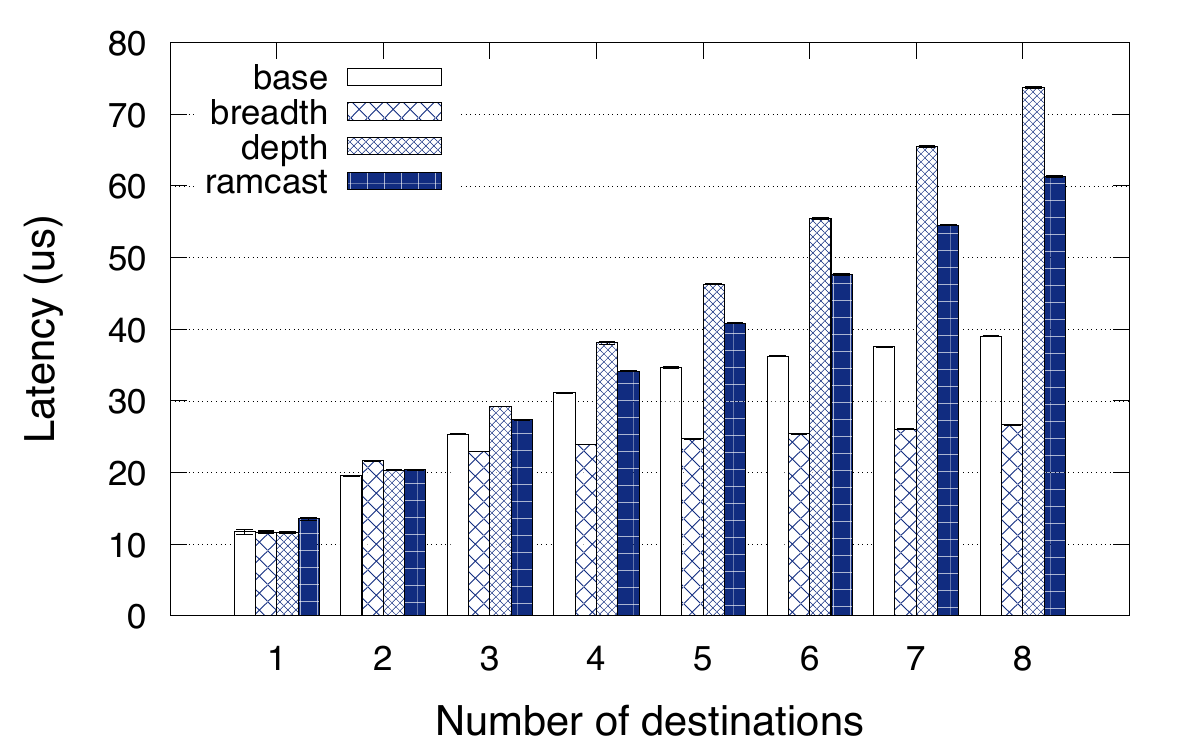} &
    \includegraphics[width=\mygraphsize\textwidth]{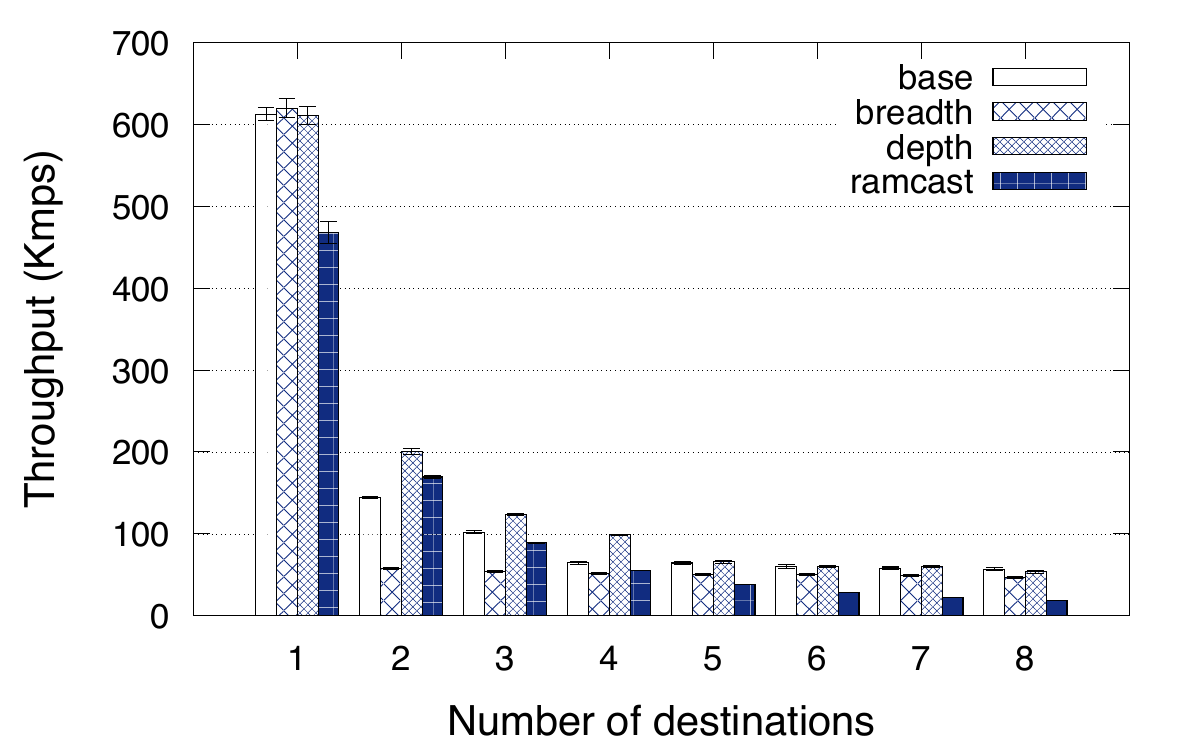} &
    \includegraphics[width=\mygraphsize\textwidth]{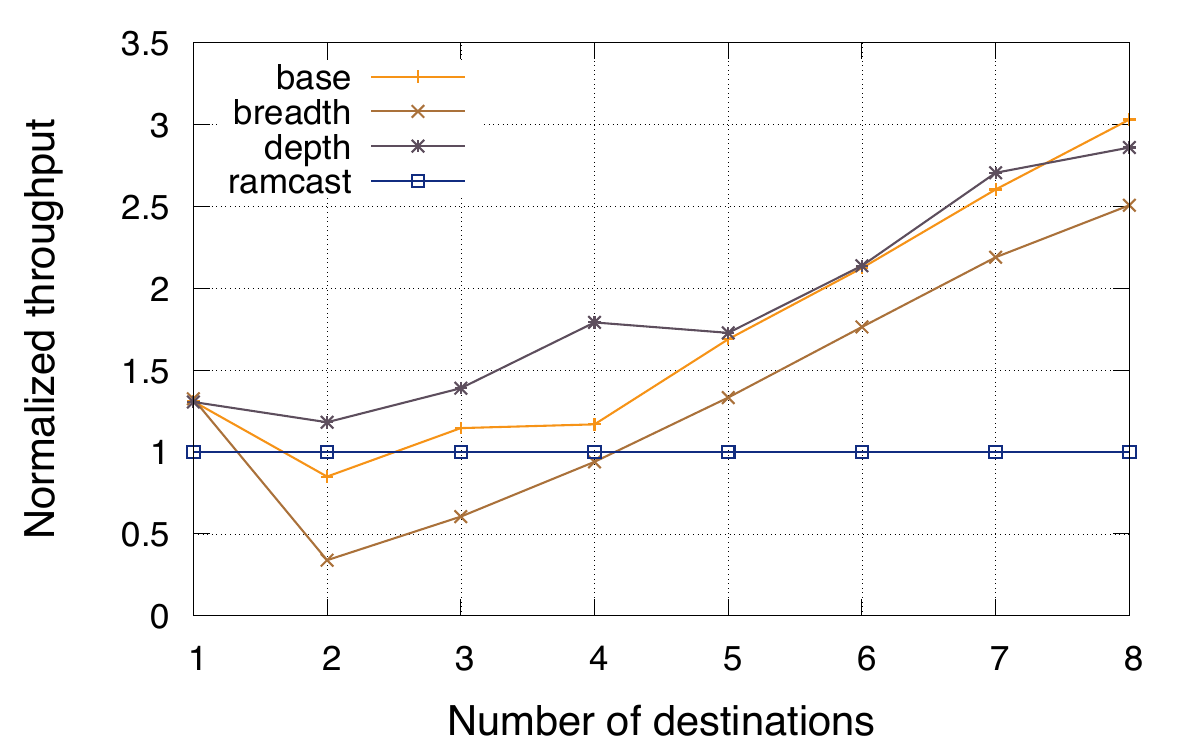} \\
  \end{tabular}
  \caption{Latency (top), throughput (center), and normalized throughput wrt RamCast (bottom) of \libname and RamCast under a ``genuine'' workload with 8 groups (3 replicas each) and 64B payload. Latency is measured with one client; throughput at system saturation.}
  \label{fig:perf:skew}
\end{figure}

Figure \ref{fig:perf:skew} considers a workload in which messages are multicast to adjacent groups in the tree. This represents an ideal scenario for \libname, as all groups involved in ordering a message are message destinations (so even \libname, as RamCast, behaves as a genuine protocol), and in fact we measure better performance in comparison with the previous case.
In this scenario, \libname throughput is always better than RamCast when using the \emph{depth} tree.

\subsection{Impact of payload size}
\label{sec:perf:msize}

\begin{figure}[htbp]
  \centering
  \begin{tabular}{cc}
    \includegraphics[width=\mygraphsize\textwidth]{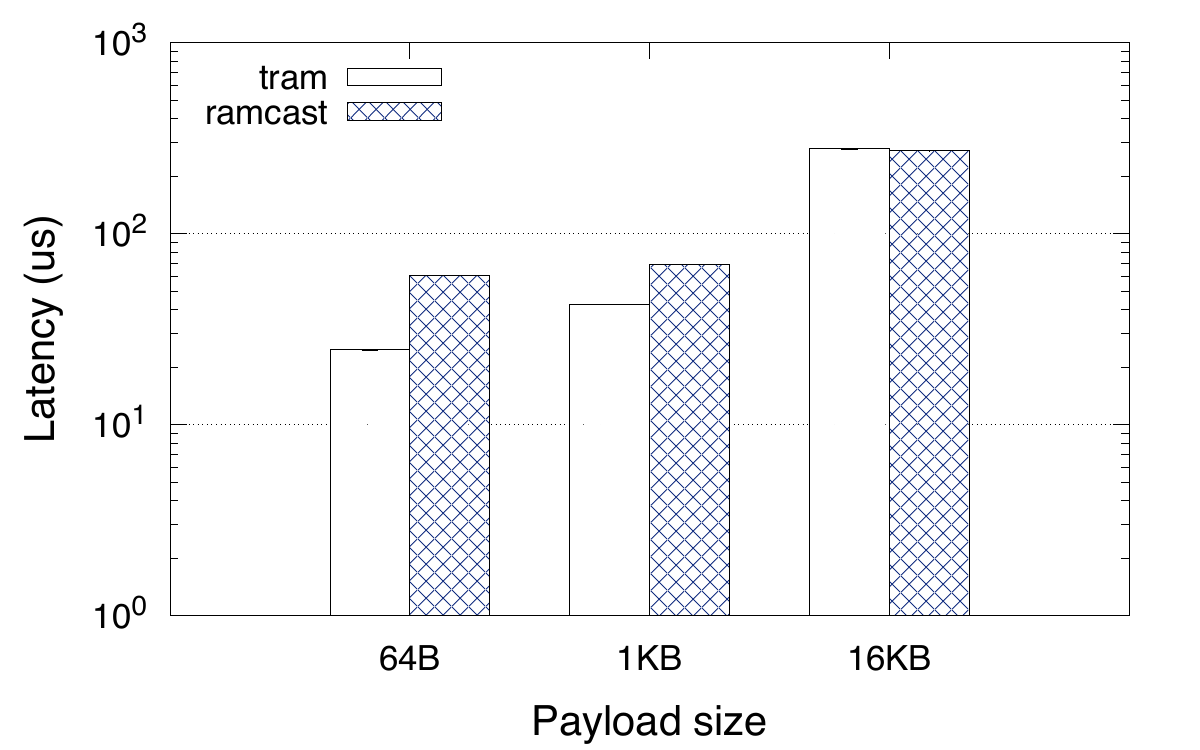} &
    \includegraphics[width=\mygraphsize\textwidth]{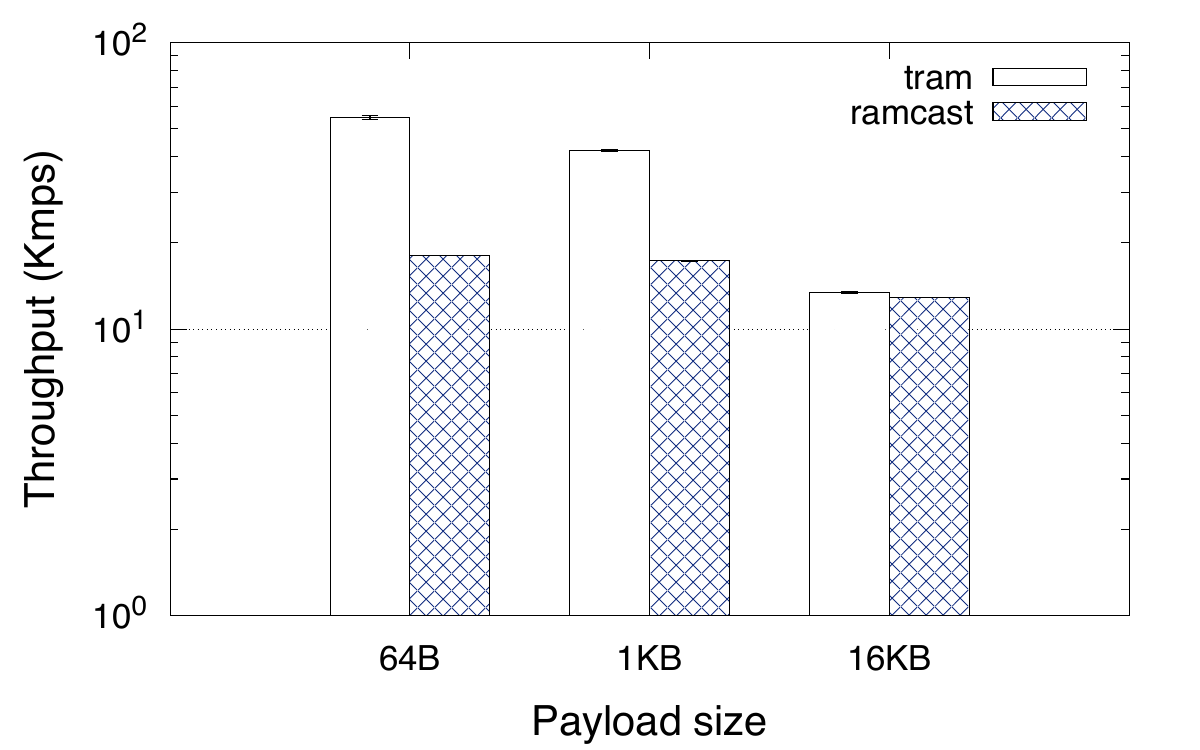}   \\
  \end{tabular}
  \caption{Comparison of latency and throughput of \libname and RamCast as the payload size increases. The experiment is run with three replicas per group and messages sent to 8 destinations. \libname uses the breadth tree configuration for latency and the depth tree configuration for throughput.}
  \label{fig:perf:msize}
\end{figure}

Figure \ref{fig:perf:msize} shows the effect of payload size on latency and throughput for \libname and RamCast.
With small payloads, \libname outperforms RamCast, but the performance gap narrows as the payload grows, converging around 16 KB where both protocols achieve similar results.
The initial advantage of \libname stems from its lower overhead.
However, \libname stores the message payload in $HM$ and $LOG$, whereas RamCast maintains only a single copy.

\subsection{Impact of replication}
\label{sec:perf:replic}

\begin{figure}[htbp]
  \centering
  \includegraphics[width=\mygraphsize\textwidth]{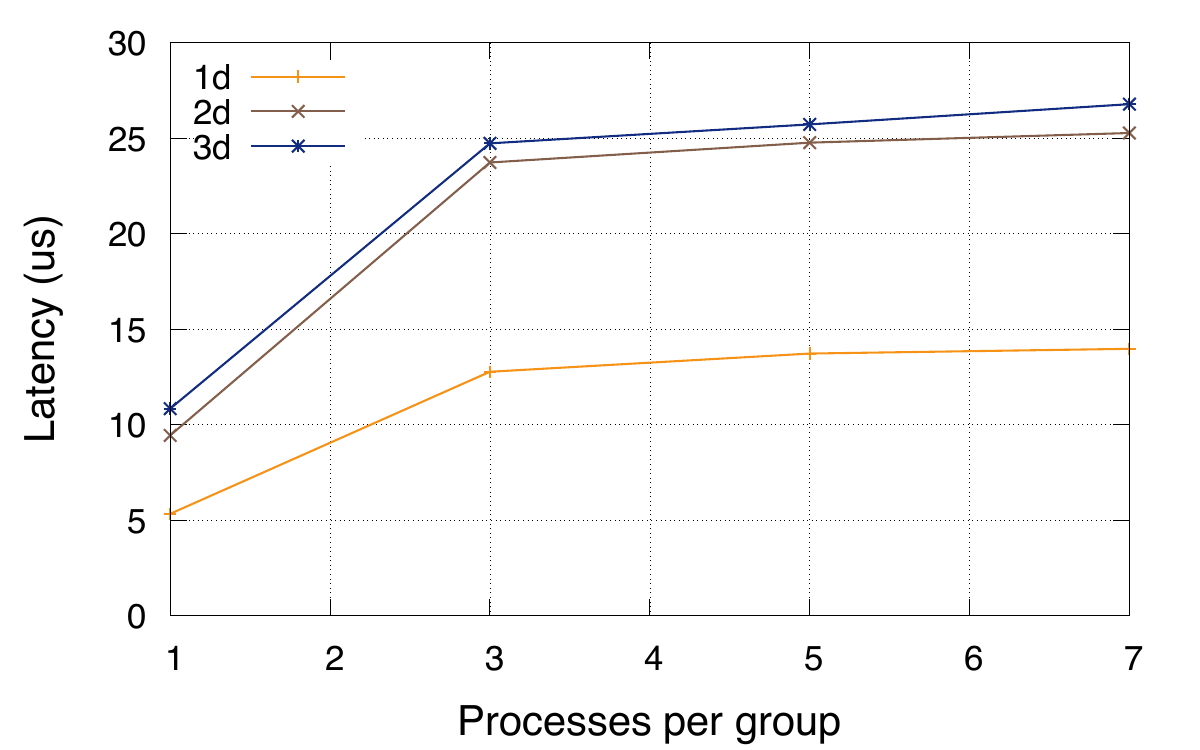}
  \caption{Latency of \libname simple tree as the number of processes per group increases, with a single client and 64B payload.}
  \label{fig:perf:replic}
\end{figure}

Figure \ref{fig:perf:replic} depicts the latency of \libname as the number of replicas per group increases. This setup deploys a simple tree with three groups (two connected to the root). As expected, the parallel writes on passive replicas allow \libname to perform replication in almost constant time, independently of the number of processes per group. This is true independently of the number of destinations.

\subsection{Shared memory versus message passing protocols}
\label{sec:perf:others}

\begin{figure}[htbp]
  \centering
  \begin{tabular}{cc}
    \includegraphics[width=\mygraphsize\textwidth]{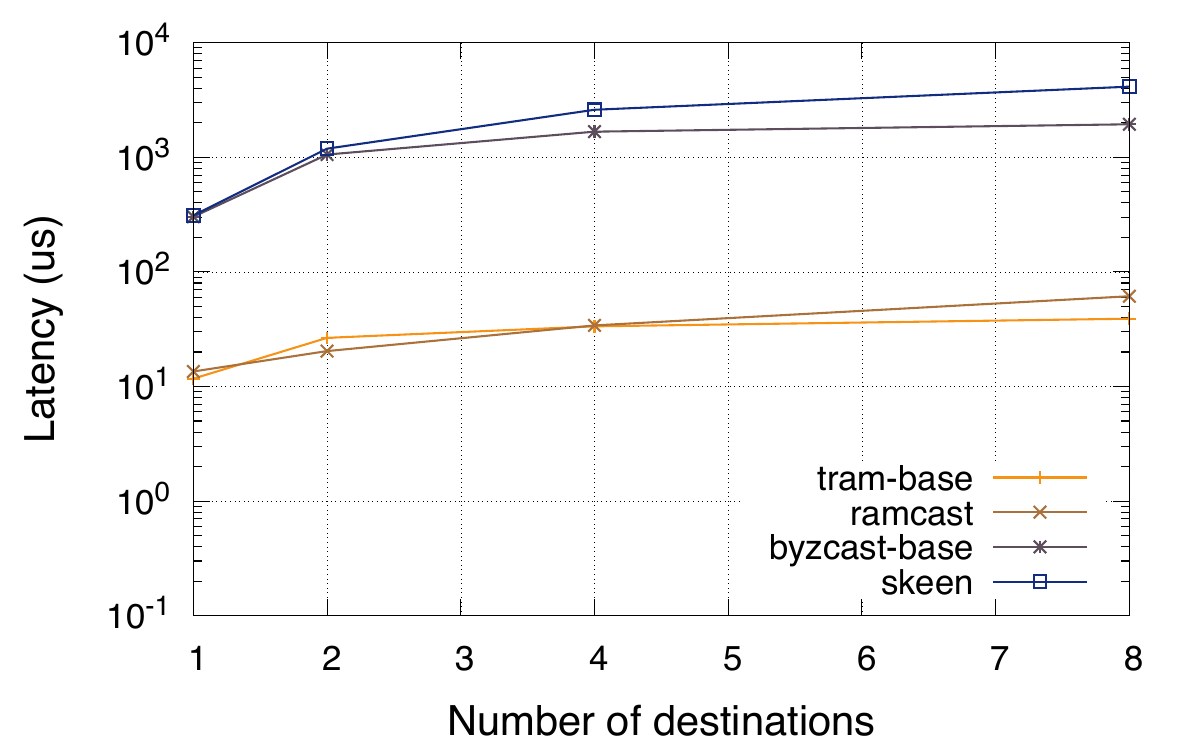} &
    \includegraphics[width=\mygraphsize\textwidth]{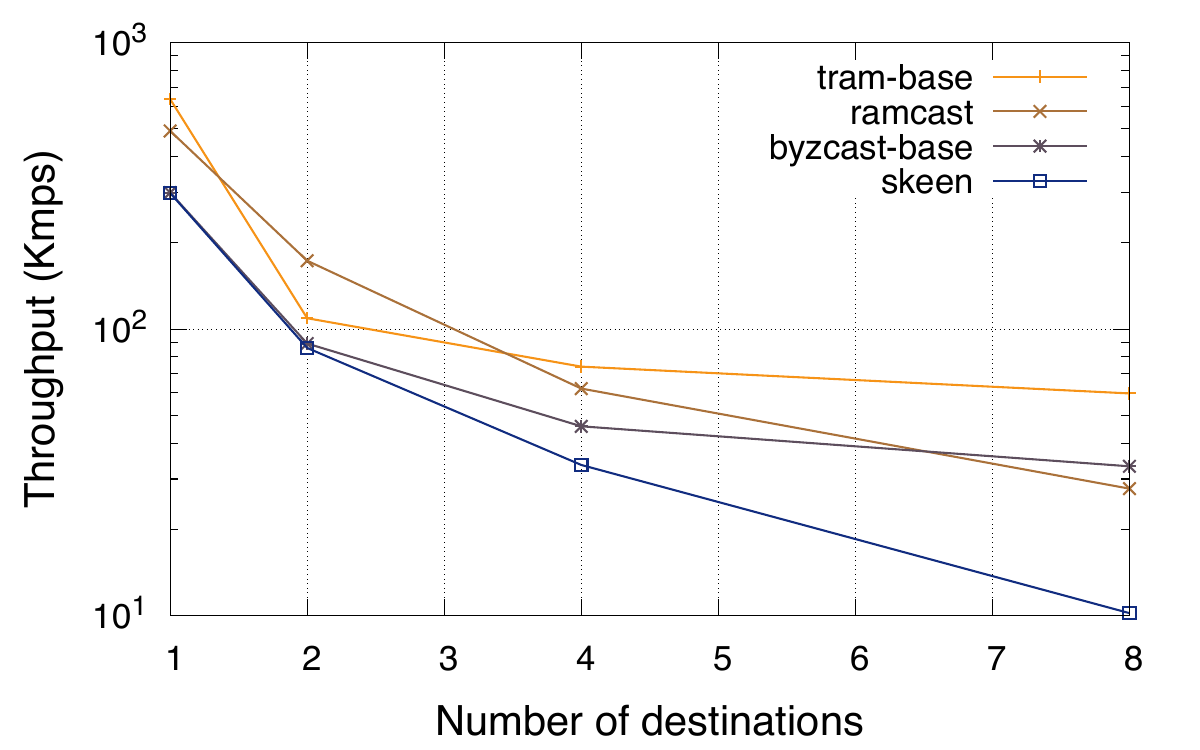}   \\
  \end{tabular}
  \caption{Latency (top) and throughput (bottom) of \libname base tree and RamCast compared to their message-passing counterparts, with a 64B payload.}
  \label{fig:perf:others}
\end{figure}

Figure \ref{fig:perf:others} compares the latency and throughput of \libname and RamCast to their respective message-passing counterparts. For \libname and ByzCast we consider the \emph{base} tree, which provides the best compromise in terms of latency and throughput. Note also that the ByzCast \cite{ByzCast} and Skeen implementations shown here are non-replicated and implemented in Java. The key observation we gather from these experiments is that, as expected, ByzCast exhibits a similar trend to \libname and Skeen parallels RamCast. More importantly, the shared memory solutions outperform the message-passing implementations by orders of magnitude.

\subsection{Impact of failures and leader changes}

In the presence of leader failures, a new leader must be found and the group/tree must be updated, which is a complex task, as described in Section~\ref{sec:algodetails}. To measure the impact of performing this complex task on \libname, we consider the \emph{breath} tree with three replicas per group and test a worst case scenario in which a leader change is triggered at the same time in each one of the 8 groups, while 100 clients send messages in parallel, each one addressed to all 8 groups. This leader change is triggered three times (waiting for the previous change to complete before triggering a new change) and Table \ref{tab:failover} shows the statistics gathered by clients over all and only the messages delayed by the leader change protocol (during each one of the three changes and in total).

\begin{table}[h!]
  \small
  \centering
  \begin{tabular}{|c|c|c|c|c|}
    \hline
    Leader change event & Average & St.Dev. & Min   & Max   \\
    \hline
    \#1                 & 22544   & 17424   & 4420  & 50730 \\
    \#2                 & 39344   & 9397    & 29879 & 48806 \\
    \#3                 & 28151   & 7038    & 21925 & 41777 \\
    \hline
    Total               & 28251   & 13430   & 4420  & 50730 \\
    \hline
  \end{tabular}
  \caption{Failover latency (us)}
  \label{tab:failover}
\end{table}

Comparing these results with those of Figure~\ref{fig:perf:design}, we notice that the latency of delayed messages can be greater than the normal case by 3 orders of magnitude. This is a result of \libname using RDMA permissions to operate. Indeed, while RDMA permissions allow to optimize protocol execution in absence of failures, they require a full reconnect of the RDMA queue pairs when a remote operation fails due to denied permissions, which is a very slow operation.

The difference in the statistics collected for the three changes (one per row in table) and the big variance we measured among the affected messages, hints at the fact that latency measured at each client strongly depends on the progress of the failover procedure at the moment the client multicasts a message.




\section{Related work}
\label{sec:related-work}


Many ordering protocols target message-passing systems, focusing on single-shard consistency. A seminal example is Paxos \cite{L98}, a fault-tolerant consensus protocol for SMR. Paxos defines three roles: proposers, who propose values; acceptors, who choose values; and learners, who learn the decided value. The Ring Paxos protocol \cite{marandi2010ring} is a high-performance variation of Paxos tailored for clustered systems. It enhances efficiency by decoupling message ordering from payload propagation, optimizing communication, and reducing overhead in distributed environments. Raft \cite{184040} is a consensus algorithm that separates key consensus elements (i.e., leader election, log replication, and safety) to simplify system design and reduce complexity. S-Paxos \cite{biely2012s} is an SMR protocol for clustered networks, derived from Paxos. It achieves high throughput by distributing the load evenly among replicas, leveraging otherwise idle resources, and mitigating the leader bottleneck typically seen in traditional Paxos implementations.

Atomic multicast protocols ensure consistent message propagation in multi-sharded systems. One of the earliest, proposed by D. Skeen~\cite{BJ87b}, used timestamps for ordering but lacked fault tolerance. Later protocols extended Skeen's approach to tolerate failures (e.g., \cite{FastCast17, fritzke1998fault, gotsman2019white, rodrigues1998scalable}). These protocols treat destinations as process groups, where SMR ensures reliability despite individual process failures~\cite{Sch90}.
Recent protocols aim to reduce the cost of replication within groups while keeping Skeen's original timestamping mechanism.
FastCast \cite{FastCast17} improves performance by optimistically executing parts of the replication logic within a group in parallel.
WhiteBox atomic multicast \cite{gotsman2019white} uses the leader-follower approach to replicate processes within groups. PrimCast~\cite{Pacheco2023} improves the leader-follower approach relying on simple quorum intersection to deliver requests in three communication delays to every replica in the destination.

Some atomic multicast protocols restrict process communication using a hierarchical overlay that determines how groups can communicate.
The protocol in \cite{DGF00} avoids using timestamps to order messages. Instead, the protocol assigns a total order to groups and processes messages sequentially through their destination groups based on this order.
Byzcast~\cite{ByzCast} is a byzantine fault-tolerant atomic multicast that arranges groups in a tree overlay.
Similarly to \libname, Byzcast is not genuine, as delivering messages to multiple groups may involve intermediary groups outside the destination set. FlexCast \cite{flexcast} is a hierarchical protocol that achieves genuineness by utilizing a complete directed acyclic graph overlay, instead of a tree, employing a history-based protocol for message ordering.
In Multi-Ring Paxos \cite{marandi2012multi}, processes subscribe to the groups they are interested in receiving messages from, and then a deterministic merge procedure is used to ensure a partial ordering of messages.

Many atomic broadcast protocols target shared memory systems. DARE \cite{DARE} minimizes latency by using RDMA one-sided operations for leader replication and permission-based leader transitions. APUS \cite{APUS} enhances DARE by combining RDMA with Paxos to improve scalability with increasing connections and replicas. It intercepts inbound socket calls, allowing seamless integration without requiring application modifications. Derecho \cite{jha2019derecho} organizes applications into subgroups and shards, supporting SMR within each. It uses a Paxos variation for updates, a novel snapshot isolation for queries, and virtual synchrony for dynamic membership. However, it lacks an abstraction to enforce total order across multiple subgroups and shards. Mu \cite{Mu} implements Protected Memory Paxos and colocates the client with the Paxos leader to achieve low latency. Mu's approach greatly simplifies the logic needed to achieve fault tolerance, since a faulty leader implies a faulty client. Despite its performance, this colocation prevents it from being applied to atomic multicast protocols.

RamCast \cite{le2021ramcast} is a shared-memory genuine atomic multicast algorithm that combines Skeen's algorithm, the leader-follower model, and RDMA technology. RamCast orders messages to a single group with 2 RDMA write delays at the leader, and to multiple groups with 3 RDMA write delays involving both the leader and followers.
P4ce \cite{P4CE} is a replication protocol that, building on previous RDMA-based consensus protocols, achieves consensus in a single round-trip. It decouples the decision-making from RDMA communication, running consensus decisions on a commodity server while a programmable switch handles the communication part.
FaRM \cite{FaRM} is a main memory distributed platform that uses RDMA to enhance latency and throughput. It exposes cluster memory as a shared address space, allowing applications to perform transactions with location transparency. FaRM boosts performance with lock-free RDMA reads and supports object colocation and function shipping for efficient single-machine transactions.
\section{Conclusions}
\label{sec:conclusions}

This paper introduces \libname, a novel tree-based atomic multicast protocol optimized for shared-memory environments on RDMA technology.
By leveraging the advantages of an overlay tree structure, \libname achieves significant performance improvements in both throughput and latency compared to existing shared-memory and message-passing-based protocols.
Our evaluations demonstrate that \libname not only outperforms state-of-the-art protocols like RamCast under uniform workloads but also adapts effectively to varying system demands.
By combining the efficiency of RDMA operations with the structured flexibility of tree-based communication, \libname addresses the challenges of fault tolerance, scalability, and consistency in distributed systems.
These results underscore the potential of \libname as a robust foundation for modern high-performance, multi-shard systems.
Future research may explore further optimizations in tree configurations and extend \libname's applicability to dynamic and heterogeneous system environments.

\bibliographystyle{unsrt}  
\bibliography{references}  

\end{document}